\documentstyle[12pt]{article}
\def\be{\begin{equation}}
\def\fe{\end{equation}}

\def\spose#1{\hbox to 0pt{#1\hss}}\def\lta{\mathrel{\spose{\lower 3pt\hbox
{$\mathchar"218$}}\raise 2.0pt\hbox{$\mathchar"13C$}}}  \def\gta{\mathrel
{\spose{\lower 3pt\hbox{$\mathchar"218$}}\raise 2.0pt\hbox{$\mathchar"13E$}}} 

\def\Libra{\spose {--} {\cal L}}

\begin{document}

\title{\bf Covariant  analysis of Newtonian multi-fluid
models for neutron stars: 
I Milne-Cartan structure and variational formulation}

\author { {\bf Brandon Carter \& Nicolas Chamel }\\
 \hskip 1 cm\\   \\Observatoire de
Paris, 92195 Meudon, France.}

\date{\it December, 2003}

\maketitle 

\vskip 1 cm
{\bf Abstract} This is the first  of a series of articles showing how 4 
dimensionally covariant analytical procedures developed in the context of 
General Relativity can be usefully adapted for application in a purely 
Newtonian framework where they provide physical insights (e.g. concerning 
helicity currents) that are not so easy to obtain by the traditional approach 
based on a 3+1 space time decomposition. After an introductory presentation 
of the relevant Milne spacetime structure and the associated Cartan 
connection, the essential principles are illustrated by application to the 
variational formulation of simple barotropic perfect fluid models. This 
variational treatment is then extended to conservative multiconstituent self
gravitating fluid models of the more general kind that is needed for
treating the effects of superfluidity in neutron stars.

\vskip 1.6 cm

\bigskip
{\bf 1. Introduction}
\medskip

As a generalisation of previous work ~\cite{CarterKhalatnikov94} 
on the special case of Landau's two-constituent superfluid model, 
using a 4-dimensionally covariant treatment of the kind pioneered
by Peradzynski~\cite{Peradzynski90}, this article presents a coherent 
fully covariant approach to the construction and application of Newtonian 
fluid models of the more general kind required in the context of neutron 
star phenomena in cases for which it is necessary to allow for independent
motion of neutronic and protonic constituents.

    Whereas a simple perfect fluid model is sufficient for deriving 
the most basic features of neutron stars (such as the radius
for a given mass and the oblateness for a given angular momentum)
models involving at least two independent constituents (of which 
one at least is superfluid) are needed to account for the details
revealed by pulsar frequency observations. If quantitative accuracy 
is needed, the high mean densities of neutron stars require the use 
of a general relativistic treatment.  In accordance with this
requirement, in applications for which a simple perfect fluid model 
is sufficient, use of fully relativistic models has been standard 
practice since the outset of neutron star theory. However, when more 
elaborate models have been needed, most work has relied on less accurate
Newtonian models, either because the relevant relativistic models had 
not been developed or because, even if available in principle, the 
relevant relativistic models were too difficult to apply in practice.

When both non relativistic and relativistic versions are available, as 
is the case~\cite{CarterKhalatnikov94,Carter89} for multiconstituent 
superfluid models, the question of which is most appropriate for a given 
purpose depends not just on considerations of intrinsic accuracy or 
computational economy but also on questions of extrinsic compatibility 
with the relevant background framework. Thus for treating perturbations 
of a zeroth order global configuration described by a fully relativistic 
perfect fluid model, what will usually be most convenient is the 
employment~\cite{LangloisSedrakianCarter98,AnderssonComer01} of a two 
constituent fluid model that is  also fully relativistic. However to 
deal with interactions with a solid crust described by a Newtonian 
elasticity model (since although appropriate relativistic elasticity 
models are available in principle~\cite{Carter89}, their technical 
complexity has so far prevented them from being effectively applied in 
practice) it may be more practical
~\cite{CarterLangloisSedrakian00,PrixComerAndersson01} 
to use a two constituent fluid model that is also non-relativistic. 

  The purpose of this article is to show how to set up and apply a 
{\it fully covariant} formulation of the kinds of non-relativistic 
multiconstituent fluid dynamical models that are needed for such cases.
Using a more traditional kind of formulation based on a preferred
space reference frame (which complicates the treatment of effects
such as helicity conservation, but facilitates the generalisation to allow 
for electromagnetism) a complementary development of the same class of 
Newtonian models has recently been provided by Prix~\cite{Prix02}.

The previous analysis on which the present work is
based~\cite{CarterKhalatnikov94} was restricted to the Landau model which 
involves just a single massive particle constituent together with a second 
constituent, representing entropy, that is massless in the Newtonian limit. 
The more general analysis presented here covers cases (including the 
historic prototype of the Andreev Bashkin model \cite{AndreevBashkin75} for
a superfluid helium mixture) involving at least two independent constituents 
representing particles of which {\it both} kinds are massive, as in the 
particularly relevant example of the application to independently moving 
superfluid neutrons and protons. One of the side issues that will be 
dealt with here is the relationship between the effect of entrainment 
(whereby the relevant momenta deviate from the corresponding particle
velocities) and the ``effective masses'' that have been defined in 
different ways in the published literature. 

As has already been pointed out by Peradzynski~\cite{Peradzynski90}, 
a noteworthy advantage (even for non-relativistic description) of using
a 4-dimensionally covariant treatment, like that of the ``canonical'' 
approach developed here,  is the possibility of exploiting Cartan type
methods of mathematical analysis, involving the use of antisymmetric 
differential forms for the construction, not just of vorticities, but 
also (as in the relativistic case~\cite{CarterKhalatnikov92}) of a more 
elaborate category of helicity currents. The technical intricacy of the 
construction procedure for these helicity currents is such that they 
would be very awkward to deal with (and are therefore usually ignored) 
in the conventional kind of approach based on a non-covariant frame 
 formulation of Newtonian theory.

   When setting up a covariant description of Newtonian theory it is 
worthwhile to recall how, having already jettisoned the distinction 
between time and space in his special relativity theory, Einstein went 
on to jettison the distinction between spacetime and gravitation in his 
general relativity theory: in the latter theory the specification of 
the spacetime metric $g_{\mu\nu}$ automatically includes the
specification of gravity via the corresponding Riemannian covariant
differentiation operator $\nabla_{\!\mu}$ which is determined by a
connection with components $\Gamma_{\!\mu\ \rho}^{\ \nu}$ that are 
given by the well known Christoffel formula $\Gamma_{\!\mu\ \rho}^{\ \nu}
$ $=g^{\nu\sigma}\big(g_{\sigma(\mu,\rho)} -{1\over2}
g_{\mu\rho,\sigma}\big)$, using a comma for partial differentiation with
respect to the (arbitrarily chosen) space time coordinates $x^\mu$, and
round brackets for index symmetrisation. However a distinction between 
spacetime and gravity can be made in the Newtonian analogue of Einstein's 
Riemannian structure, namely something that is known
~\cite{Kunzle86,DuvalGibbonsHorvathy91} as a Newton Cartan structure -- 
whose not so widely familiar principles will be recapitulated below -- 
involving a non-Riemannian covariant differentiation operator $D_\mu$ 
that is determined by a Cartan connection with components 
$\omega_{\mu\ \rho}^{\ \nu}$ that are {\it not} obtainable from a 
Christoffel formula because the Newtonian structure does not specify 
the non-degenerate spacetime metric that would be needed. In contrast 
with the inextricable case of general relativity, it is possible in the 
Newtonian case to extricate the specification of the gravitational field, 
as given by the (non-tensorial) Cartan connection components 
$\omega_{\mu\ \rho}^{\ \nu}$, from that of the underlying spacetime 
manifold. Prior to any knowledge of the gravitational field (i.e. the 
Cartan connection) the only endowment of the underlying Newtonian 
spacetime manifold consists just of what is describable as a Milne 
structure, a concept that will be briefly recapitulated in the 
immediately following section. 

    Like the Minkowski structure of special relativity theory, the Milne 
structure of Newtonian spacetime does not involve any free parameters or 
fields, i.e. it is intrinsically unique (modulo diffeomorphisms). 
Nevertheless, despite its intrinsic simplicity the nature of this 
Newtonian spacetime structure is just sufficiently subtle to have
prevented it from being properly understood until Milne's introduction 
of Newtonian cosmology theory a couple of decades {\it after} Einstein's 
introduction of general relativity --though not so long after 
Friedmann's foundation of the corresponding general relativistic 
cosmology theory -- at about the same time as Cartan's epoch making work 
on the development of the appropriate mathematical machinery. Milne's 
breakthrough was based on the extrapolation to a global level of 
Einstein's earlier observation -- originally in a Newtonian context, as
a guiding principle for the construction of the corresponding
relativistic theory -- of the equivalence at a local level between 
gravitation and acceleration.

\vfill\eject
\bigskip
 {\bf 2. Covariant description of Newtonian spacetime}
\medskip

   Whereas a fully covariant treatment is generally recognised to be 
indispensable for formulating general relativistic models, on the other 
hand, for their Newtonian analogues, the usual practice is to rely entirely 
on an `Aristotelian' decomposition whereby space time is considered as
a direct product of a flat Euclidean 3-space with a one dimensional
Euclidean time line. Any such {\it Aristotelian structure} will be
characterised by a corresponding class of Aristotelian coordinate systems,
which consist of a set of ordinary Cartesian (orthonormal) space 
coordinates $X^i$ ($i=1,2,3$) together with a Newtonian time coordinate
$t$ which is physically well defined modulo a choice of time origin.
These coordinate systems are mapped onto each other by the transformations
of a 7 parameter Aristotelian symmetry group, consisting of the product of 
the 6 parameter group of Euclidean translations and rotations with the
1 parameter group of time translations. 

Whereas the time coordinate $t$ of such a system is physically well
defined (modulo an arbitrary adjustment of the origin) it has been
generally recognised since the foundation of Newtonian theory that in 
a generic application there will be no uniquely preferred Aristotelian 
structure, but that the theory will be invariant with respect to a group 
of {\it gauge transformations} relating different Aristotelian product 
structures that all share the same constant time sections but that do not 
have the same sections of constant space position (as measured by fixed 
values of the Aristotelian space coordinates $X^i$). Any such gauge 
transformation will be specifiable by a mapping of one of the (7
parameter family of) sets of Aristotelian coordinates of a particular 
Aristotelian structure to a set belonging to another such structure. Since 
the flat constant time sections are preserved, any such transformation 
must be expressible just as a  mapping $X^i\mapsto\breve X^i$, 
$t\mapsto\breve t$, for which the time transformation is trivial,
$\breve t=t$, and for which the new space coordinates $\breve X^i$ are
given by a time dependent Euclidean transformation. However not all kinds 
of time dependent Euclidean transformation are admissible. In particular 
time dependent rotations (belonging to what is known as the Coriolis group) 
are excluded from the status of gauge transformations because they change 
the physical comportment of the system (giving rise to what is known
as the Coriolis effect). 

It turns out that the only admissible gauge transformations between 
different Aristotelian structures are time dependent space translations as 
given by a transformation of the form $X^i\mapsto \breve X^i$ with
\be \breve X^i= X^i- c^i\, ,\label{09}\fe
for quantities $c^i$ that are constants in the sense that they have to be 
independent of the space coordinates $X^i$, but that are arbitrarily 
variable as functions of time. For a long time it was generally believed 
that not all such time dependent translations were admissible, but only the
three parameter set of Galilean transformations, meaning those that are 
{\it linear} so that the quantities $c^i$ can be taken to be given by 
expressions of the form $c^i=b^i t$ in which the quantities $b^i$ are 
constants in the strong sense of being independent not just of the $X^i$ 
but also of $t$. A set of Aristotelian structures related by such linear 
transformations constitutes what may be described as a Galilei structure. 

The important point that escaped everybody's notice until, after having
been implicitly recognised in Einstein's ``equivalence principle'', it was 
finally exploited for the foundation of Newtonian cosmology by 
Milne~\cite{Milne34} is that Newtonian mechanics contains nothing to 
distinguish any particular preferred Galilei structure. The set of gauge 
transformations that are admissible in the sense that (unlike generic 
Coriolis transformations) they preserve the form of the physical laws 
of motion is not restricted to linear Galilean transformations, but 
includes generic transformations of the form (\ref{09}) in which, while 
independent of the $X^i$, the quantities $c^i$ are allowed to have an 
arbitrarily non-linear dependence on the time $t$. The complete set of all 
the Aristotelian product decompositions that can be obtained by such Milne 
gauge transformations constitutes what may be described as the {\it Milne 
structure} of space time. Thus the (physically unique) Milne structure 
consists of a family of (physically equivalent) Galilean structures that 
are related to each other by {\it accelerated} space translation 
transformations of the form (\ref{09}), while each member of the (infinite
parameter set) of Galilei structures consists of a (3 parameter) set of 
Aristotelian structures that are related to each other by {\it linear} 
transformations of the form (\ref{09}).

Although the traditional employment of a particular choice of Aristotelian 
structure, and in particular of some corresponding set of Aristotelian 
coordinates $\{t,X^i\}$ is useful for many purposes, such as the 
exploitation of the flatness and distant parallelism of the preferred
(constant time) 3-space sections, the advantages of this are  not cost free. 
The price to be paid includes not just the well known obligation to verify 
Galilean (and Milne) invariance with respect to changes of the 
Aristotelian frame. Another, less well known, cost is the loss of access 
to the elegant and powerful mathematical methods  based on tensors, and 
particularly on Cartan type differential forms, that become available 
when a fully covariant four dimensional framework is used.

A convenient feature of the  3 dimensional constant time sections in an 
Aristotelian decomposition is the existence of a physically well defined - 
and flat - metric having components $\gamma_{ij}$ that are given simply by 
the unit matrix with respect to orthonormal Aristotelian space coordinates 
$X^i$ ($i,j=1,2,3$). This metric can be used together with its contravariant 
inverse $\gamma^{ij}$ for raising and lowering of space indices, including 
those of the associated antisymmetric volume measure tensor 
$\varepsilon_{ijk}$ (whose non vanishing components are given by $\pm 
\sqrt{{\rm det} \{\gamma\}}$ with the sign depending on whether the 
$\{i,j,k\}$ is an even or odd permutation of $\{1,2,3\}$) whose 
contravariant version will be characterised by the normalisation condition 
$\varepsilon^{ijk} \varepsilon_{ijk}=3!$. A similarly convenient feature of 
a general relativistic formulation, is the existence of a physically well 
defined - but not in general  flat - space time metric giving components 
$g_{\mu\nu}$ say with respect to coordinates $x^\mu$ ($\mu,\nu$ = 0,1,2,3), 
which determines a corresponding antisymmetric spacetime volume measure 
tensor with non vanishing components equal to $\pm\sqrt{-{\rm det}\{g\}}$, 
and that can be used together with its contravariant inverse $g^{\mu\nu}$ 
for raising and lowering of space time indices. Quite apart from the 
seductive flatness and parallelism properties of the preferred space
sections in an Aristotelian decomposition, one of the reasons why 
fully covariant formulations of Newtonian dynamics are not as widely used as 
they deserve to be is their lack of an analogous means of raising and
lowering of space time indices. One of the purposes of this work is to show 
that, despite this handicap, Newtonian mechanics can nevertheless be set up 
without too much difficulty in a fully covariant formulation that makes it 
easy to exploit the technical advantages of freedom to use arbitrarily 
curved space time coordinates $x^\mu$.

The first step is to obtain the fundamental spacetime tensor fields that 
are available, in lieu of the relativistic metric tensor, for the
characterization of Newtonian space time. To start with, for any fixed 
value of the preferred Newtonian time coordinate, $t$, the embedding 
mapping $X^i\mapsto x^\mu$ of the corresponding Aristotelian space section 
will determine physically well defined contravariant spacetime tensor 
fields with components given by
\be\gamma^{\mu\nu}=\gamma^{ij}x^\mu_{\, ,i} x^\nu_{\, ,j}\, \hskip 1 cm  
\varepsilon^{\mu\nu\rho}=\varepsilon^{ijk} x^\mu_{\, ,i} x^\nu_{\, ,j} 
x^\rho_{\, ,k}\, ,\label{00}\fe
using a comma to indicate partial differentiation (in this case with 
respect to the space coordinates $X^i$). However there is no 
unambiguously preferred (Galilei invariant) prescription for lowering 
the indices to obtain corresponding contravariant versions 
$\gamma_{\mu\nu}$ and $\varepsilon_{\mu\nu\rho}$ because the former is 
degenerate ($\gamma^{\mu\nu}$ has matrix rank 3 not 4) so it does not 
have a well defined inverse. 

       The foregoing consideration means that in Newtonian theory 
4-dimensional  tensor indices will in general have an irrevocably 
covariant or contravariant nature. The most basic example is that of 
the preferred covariant unit vector $t_\mu$ that is obtained simply as
the gradient of the preferred Newtonian time coordinate, i.e. $t_\mu$ 
$=t_{,\mu}$, and that is a null eigencovector of the degenerate 
preferred contravariant space metric, and also orthogonal to 
$\varepsilon^{\mu\nu\rho}$, i.e. 
\be \gamma^{\mu\nu}t_\nu=0\, ,\hskip 
1 cm \varepsilon^{\mu\nu\rho} t_\rho=0\, .\label{01}\fe
An exception, having both a covariant and a contravariant version, is 
that of the antisymmetric spacetime volume measure tensor 
$\varepsilon_{\mu\nu\rho\sigma}$, which has a physically well defined 
normalisation -- despite the lack of a non degenerate space time metric
in the Newtonian framework  -- that is specified by the relation
\be t_\mu={1\over 3!}\varepsilon_{\mu\nu\rho\sigma}
\varepsilon^{\nu\rho\sigma}\, ,\label{01a} \fe
and for which a corresponding contravariant version is unambiguously
definable by a normalisation condition of the same form,
\be \varepsilon_{\mu\nu\rho\sigma}\varepsilon^{\mu\nu\rho\sigma}
= -4!\, , \label{01b}\fe
as holds in general relativity, which means that it will satisfy
\be \varepsilon^{\mu\nu\rho\sigma}t_\sigma=\varepsilon^{\mu\nu\rho}
\, .\label{04}\fe

The fields $\gamma^{\mu\nu}$ and $t_\mu$ (the degenerate residue
representing all of the algebraic structure that remains from the 
relativistic spacetime metric in the Newtonian limit) constitute what 
may be termed a Coriolis structure, since it contains nothing that 
distinguishes rotating from non rotating frames. To incorporate this 
distinction, and thereby complete the covariant specification of 
Newtonian space time, we must consider what to use in place of the 
Riemannian connection and the associated covariant differentiation 
operator that in general relativity is uniquely specified by the 
spacetime metric $g_{\mu\nu}$. In the Newtonian case, a corresponding 
physically well defined but non-Riemannian connection 
$\omega_{\mu\ \rho}^{\ \nu}$ and associated covariant differentiation 
operator $D_\mu$ will be provided by the Newton-Cartan structure that is 
determined by the gravitational field in the manner to be described at the 
end of the next section.  However the only features of the Newtonian space 
time background that are well defined a priori (in the absence of 
information about the gravitational field that specifies the Newton
-Cartan structure) are those provided by the Milne structure that 
was described at the beginning of this section. 

The Milne structure by itself (without reference to the gravitational 
field) does not provide enough information for the unambiguous 
specification of a connection. What it is capable of providing is a 
connection that is gauge dependent. Specifically for each choice of gauge, 
i.e. for each choice of a particular Aristotelian product structure, 
there will be a corresponding natural connection with components  
$\Gamma_{\!\mu\ \rho}^{\ \nu}$ defined by the condition that they simply
vanish when evaluated with respect to a corresponding system
of Aristotelian coordinates $\{t,X^i\}$ (though not of course if
the orthonormal space coordinates $X^i$ were replaced by coordinates
of some other, e.g. spherical, kind). This means that with respect
to coordinates of this particular type (but not for a system of some
other, e.g. spherical, kind) the corresponding covariant differentiation
operator $\nabla_{\!\mu}$ will be identifiable with the simple
partial differentiation operator $\partial_\mu$.

The gauge dependent connection $\Gamma_{\!\mu\ \rho}^{\ \nu}$ that is 
defined in this way has two convenient properties. Firstly, since it is 
identifiable with partial differentiation in the relevant Aristotelian 
coordinate system, it is clear that since $\partial_\mu\partial_\nu = 
\partial_\nu\partial_\mu$  the covariant derivative will automatically 
inherit the analogous commutation property
\be \nabla_{\!\mu}\nabla_{\!\nu}=\nabla_{\!\nu}\nabla_{\mu}
\, ,\label{05}\fe 
i.e. the connection  $\Gamma_{\!\mu\ \rho}^{\ \nu}$ has the property
that (unlike the curved Newton - Cartan connection described below) it
is {\it flat}. The other convenient property is that although (again
unlike the gravitational field dependent Newton-Cartan connection 
$\omega_{\mu\ \rho}^{\ \nu}$) it is gauge dependent, its gauge dependence
is of a rather weak kind. Since connection components are unaffected
by linear transformations, it follows that $\Gamma_{\!\mu\ \rho}^{\ \nu}$ 
will not be affected by gauge transformations of the restricted 
Galilean type. Choosing a particular Galilean structure (i.e. a linearly 
related subclass of Aristotelian structures) is equivalent to choosing a 
particular connection of this flat type. It will be shown below how such a 
connection is affected by Milne gauge transformations of the more general 
accelerated type that relate distinct Galilean structures.

According to the preceeding definition, the Milne structure is an 
equivalence class of Aristotelian (direct product of time and flat space)
structures that are related to each other by a gauge group consisting
of time dependent space translations. In the modern (post Cartan)
technical language of fibre bundle theory this definition can be 
succinctly reformulated as follows.

In formal mathematical terms,  the Milne structure of Newtonian spacetime 
is formally describable as bundle of 3-dimensional Euclidean space fibres 
(each characterized by its own flat metric $\gamma_{ij}$) over a base 
consisting of a line endowed with a physically preferred time measure 
(specified by the coordinate $t$) for which the relevant (Abelian) 
bundle group consists just of the 3 - parameter set of Euclidean space
translations (but not rotations), which are expressible in terms of 
Cartesian coordinate $X^i$ on the fibre simply by transformations of the 
form (\ref{09}). (The exclusion of Euclidean rotations from the bundle 
group is what distinguishes the Milne structure from the more primitive 
Coriolis structure given just by the specification of the fields $t_\mu$ 
and $\gamma^{\mu\nu}$.)

Any particular section of this Milne bundle, i.e. a representation as 
a direct product of the base times the fibre, will be interpretable as 
a particular choice of an Aristotelian structure. In such a direct
product structure, the preferred time coordinate $t$ on the base and 
a choice of Cartesian coordinates $X^i$ on a space section (i.e. 
coordinates such that metric components $\gamma_{ij}$ form a unit 
matrix) will determine a corresponding set of Aristotelian spacetime 
coordinates $x^\mu$ according to the obvious specification $x^{_0}=t$, 
$x^i=X^i$. A corresponding connection is thereby definable as the one 
with respect to which covariant differentiation reduces to partial 
differentiation, i.e. the one for which $\nabla_{\!\mu}= \partial_\mu$ 
and $\Gamma_{\!\mu\ \nu}^{\ \rho}=0$, in these particular coordinates. 
(This connection could be interpreted as the Riemannian connection 
provided by a flat  Unruh type spacetime metric of the form 
$ds^2=\gamma_{ij}\,dX^i\,dX^j -C^2 dt^2$ in which $C$ could be any 
arbitrarily chosen constant speed, which might be that of light, 
but which in applications to perturbations in an asymptotically
uniform fluid background could more usefully~\cite{Unruh95} be taken 
to be the relevant sound speed.) 

\bigskip
{\bf 3. Galilean and Milne type gauge transformations}
\medskip

Although the dynamical equations of ordinary Maxwellian electromagnetic
theory are expressible in terms just of gauge independent quantities
such as the electric and magnetic fields, it is useful for many purposes, 
and indispensible for a variational formulation, to employ various gauge 
dependent entities (starting with the vector potential $A_\mu$). In the
present context (of ordinary Newtonian dynamical theory) it is analogously 
useful for many purposes, and indispensible for a variational formulation,
to employ entities whose specification depends on a particular choice of 
gauge, where in this context a ``choice of gauge'' is to be interpreted as
meaning a particular choice of Aristotelian structure within the large 
equivalence class of Aristotelian structures that collectively constitute 
the Milne structure described above.

In terms of the Aristotelian coordinates $\{t,X^i\}$, the transformation 
to the analogous coordinates for any alternative bundle section, i.e. any 
alternative Aristotelian structure, will be expressible by a transformation 
in which the base coordinate is held fixed, i.e. $t\mapsto t$ while the 
Cartesian fibre coordinates $X^i$ are transformed according to a relation 
of the form (\ref{09}) in which the translation vector $c^i$ is given as an 
arbitrarily variable function of the base coordinate $t$. It is evident that 
the connection specified by the new gauge will be identical with the one 
specified by the old section, i.e. we shall have 
$\breve\Gamma_{\!\mu\ \rho}^{\ \nu}= \Gamma_{\!\mu\ \rho}^{\ \nu}$
and hence $\breve\nabla_{\!\mu}= \nabla_{\!\mu}$, so long as the 
transformation (\ref{09}) is {\it linear}, i.e. so long as the translation 
is of the Galilean form characterised by the condition that the components 
\be b^i= {dc^i\over dt}\, , \hskip 1 cm b^{_0}=0\, , \label{11a} \fe
of the boost velocity vector of the transformation should be fixed, 
independently of $t$. However the connection will not be preserved if there 
is a non-vanishing value for the corresponding relative acceleration vector 
$a^\mu$ as given by
\be a^i={db^i\over dt}\, ,\hskip 1 cm a^{_0}=0 \, .\label{11b}\fe

It can be verified that for the new section obtained by a generic (i.e. 
accelerated) Milne gauge transformation the corresponding new connection 
will be related to the original one by a relation of the simple but non 
trivial form $\Gamma_{\!\mu\ \rho}^{\ \nu}\mapsto 
\breve\Gamma_{\!\mu\ \rho}^{\ \nu}$ with
\be \breve\Gamma_{\!\mu\ \rho}^{\ \nu}=\Gamma_{\!\mu\ \rho}^{\ \nu}
-t_\mu a^\nu t_\rho\, ,\label{06}\fe
where $a^\nu$ is the relevant transformation generator, which can be any 
vector that is spacelike and spacially uniform  (i.e. purely time dependent) 
in the sense that
\be t_\mu a^\mu=0\, ,\hskip 1 cm \gamma^{\mu\nu}\nabla_{\!\nu}a^\rho
=0\, .\label{07}\fe            
This transformation law has the noteworthy feature of preserving the trace 
of the connection, i.e. it gives
\be\breve\Gamma_{\!\mu\ \nu}^{\ \nu}= \Gamma_{\!\mu\ \nu}^{\ \nu}
\, .\label{08}\fe
It follows that if $n^\mu$ is a physically well defined current 4-vector of 
the kind to be discussed in the next section then the divergence given by 
the expression $\nabla_{\!\nu} n^\nu$ will also be physically well defined 
as a gauge independent scalar field (which will vanish in the particular 
case of a current that is conserved).  

It will be convenient for future reference to introduce a scalar {\it boost 
potential} function $\beta$ that is defined, modulo an arbitrary time
dependent constant of integration $\beta\{0\}$ by,
\be b^\mu=\gamma^{\mu\nu}\nabla_{\!\nu}\beta \, ,\hskip 1 cm
\gamma^{\mu\nu}\nabla_{\!\nu} b^\rho=0 \, ,\label{19}\fe
so that in the original Aristotelian coordinate system it will be given
explicitly by an expression of the form
\be\beta=\beta\{0\}+\gamma_{ij}b^i X^j\, ,\hskip 1 cm
\gamma^{\mu\nu}\nabla_{\!\nu}\beta\{0\}=0 \, .\label{19c}\fe
It then follows that the relative acceleration vector will be given by
\be a^\mu=e^\nu\nabla_{\!\nu}b^\mu=\gamma^{\mu\nu}\nabla_{\!\nu\,}\alpha
\, ,\hskip 1 cm \alpha= e^\nu\nabla_{\!\nu}\beta\, ,\label{19a}\fe
where $e^\mu$  is the relevant ``ether'' 4-velocity vector, i.e. the unit
time lapse vector of the Aristotelian rest frame, whose components with
respect to the corresponding coordinates $\{t,X^i\}$ will be given by
$e^{_0}=1$, $e^i=0$, so that it will satisfy the conditions
\be e^\mu t_\mu=1 \, ,\hskip 1 cm \nabla_{_\mu} e^\nu=0 \, .\label{19b}\fe

Whichever bundle section may have been used to specify it in the first
place, the connection will automatically be such as to preserve the
fundamental tensors $\gamma^{\mu\nu}$ and $t_\mu$ of the (Coriolis)
spacetime structure, i.e. it will satisfy
\be \nabla_{\!\mu} \gamma^{\nu\rho}=0\, ,\hskip 1 cm
\nabla_{\!\mu}t_{\nu}=0\, ,\label{20} \fe
and hence also
\be \nabla_{\!\mu}\varepsilon^{\nu\rho\sigma\tau}=0\, .\label{20a}\fe
Conversely the specification of any particular connection that satisfies
these preservation conditions  will characterize what may be described
as the corresponding Galilei structure, which can be conceived as an
equivalence class of Aristotelian (i.e. direct product) structures
related by linear gauge transformations of the form (\ref{09}) with
vanishing value of the relative acceleration vector defined by
(\ref{11b}).

        For centuries after Newton's original development of his
theory it was taken for granted that the relevant equations of motion 
singled out a preferred Galilei structure with a corresponding
Galilean transformation group with respect to which their form remained 
covariant. What Milne realised~\cite{Milne34} was that except in the 
case of a localised system in an asymptotically empty background (such as 
the solar system example to which the early successes of Newton's theory 
were restricted) the equivalence principle prevents the prescription of 
any natural rule for preferring some particular Galilean structure
rather than another.  Thus, as a consequence of the applicability of the
equivalence principle, it transpires that the relevant equations of motion 
are covariant not just with respect to the Galilei group but also with 
respect to the larger Milne group, which relates distinct Galilei 
structures by transformations characterized by non-vanishing values of 
the relative acceleration vector $a^\mu$.

\bigskip
{\bf 4. Gravity, particle dynamics, and the Newton - Cartan connection}
\medskip

The way the foregoing considerations apply to the most basic of the 
Newtonian equations of motion, namely the equation of motion for a free 
particle in some given gravitational field, is as follows. Having specified 
the worldline of the particle giving the relevant spacetime coordinates 
$x^{\,\mu}$ as functions of the Newtonian time $t$, one can go on to define 
the corresponding 4-velocity vector defined as the time parametrised 
tangent vector given by
\be u^\mu={dx^\mu\over dt}\, ,\label{31}\fe 
of which only three components are actually independent since the definition 
automatically ensures that it satisfies the unit normalisation condition
\be u^\mu t_\mu =1\, .\label{31a}\fe
The equation of motion will then be expressible covariantly as
\be u^\nu\nabla_{\!\nu}u^\mu = g^\mu\, ,\label{32}\fe
where $g^\mu$ is the relevant gravitational field vector, which must be 
strictly spacelike to avoid inconsistency with (\ref{31a}), and which, 
more specifically, is  required to be derivable as the space gradient of a 
Newtonian potential $\phi$, i.e.
\be t_\mu g^\mu=0\, ,\hskip 1 cm g^\mu=-\gamma^{\mu\nu}\nabla_\nu\phi
\, .\label{32a}\fe

Although it was traditionally expressed in a non-covariant mathematical 
form, the relation (\ref{32}) was recognised from the outset to be
physically invariant with respect to Galilean transformations, i.e. the 
linear transformations that preserve the connection involved in the 
covariant differentiation operator $\nabla_{\!\nu}$. The crucial point 
that eluded Newton and everyone else before the time of Einstein, 
Friedmann, and Milne is that $g^\mu$ is akin to the scalar potential 
$\phi$ (and to Maxwell's covector potential $A_\mu$ as contrasted with 
the tensorially well defined electromagnetic field $F_{\mu\nu}$) in that 
it cannot be considered to be an absolutely well defined locally 
measurable vector field but should be recognised to be gauge dependent. 
It can be seen from (\ref{06}) that the relation (\ref{32}) is in fact 
invariant not just with respect to Galilean transformations but to 
generic Milne transformations as characterised by a non vanishing 
acceleration vector $a^\mu$ in (\ref{11b}), provided it is understood 
that $g^\mu$ undergoes a corresponding Milne gauge transformation of the 
simple form  $ g^\mu \mapsto\breve g^\mu$ with
\be \breve g^\mu  = g^\mu - a^\mu\, .\label{32b}\fe
This evidently entails the requirement that the Newtonian potential should 
transform according to a law of the form $\phi\mapsto\breve\phi$ with
\be\breve \phi=\phi+\alpha \, ,
\label{32c}\fe
in which scalar field $\alpha$ will be given by (\ref{19a}) as the ether 
(i.e. Aristotelian) frame time derivative of the boost potential $\beta$ . 
The freedom to adjust the specification (\ref {19c}) of the latter by 
freely choosing the time dependence of the value $\beta\{0\}$ of 
the boost potential at the Aristotelian coordinate origin corresponds 
to the calibration freedom in the specification of $\phi$ by (\ref{32a}). 
Thus, even in a fixed Aristotelian frame, the potential $\phi$ will 
be subject to trivial gauge transformations consisting of the addition of 
a purely time dependent quantity that is identifiable simply as the time 
derivative of $\beta\{0\}$.

         The idea of the Newton Cartan formulation
~\cite{Kunzle86,DuvalGibbonsHorvathy91} is to replace the gauge dependent 
differential operator $\nabla_\mu$ by a corresponding gauge covariant 
differential operator $D_\mu$ that is specified by replacing the flat 
connection $\Gamma_{\!\mu\ \rho}^{\ \nu}$ by a gravitationally modified 
connection $\omega_{\mu\ \rho}^{\ \, \nu}$ say that is given by
\be\omega_{\mu\ \rho}^{\ \, \nu}=\Gamma_{\!\mu\ \rho}^{\ \nu}
-t_\mu g^\nu t_\rho\, .\label{33}\fe
Using (\ref{32b}) in conjunction with (\ref{06}), it can be verified that 
this modified connection has the desired gauge invariance property  
$\omega_{\mu\ \rho}^{\ \, \nu}\mapsto \breve\omega_{\mu\ \rho}^{\ \, \nu}$ 
with
\be \breve\omega_{\mu\ \rho}^{\ \, \nu} =
\omega_{\mu\ \rho}^{\ \, \nu}\, .\label {33b}\fe
This  makes it possible to rewrite the dynamical equation (\ref{32}) in 
the manifestly gauge invariant (as well as coordinate covariant) form
\be u^\nu D_\nu u^\mu=0\, .\label{33a}\fe
While facilitating the comparison with general relativity, this gauge 
covariant differentiation operator $D_\mu$ has the disadvantage of lacking 
the convenient flatness property (\ref{05}) of the gauge dependent 
alternative $\nabla_{\!\mu}$. For example if we consider not just a single 
particle trajectory but a fluid flow characterized by a velocity 4-vector 
$u^\mu$ that is defined as a field over spacetime then we shall have
\be D_{[\mu}D_{\nu]}u^\rho={_1\over^2} 
R_{\mu\nu\ \sigma}^{\ \ \, \rho} u^\sigma\, ,\label{34}\fe
using square brackets to indicate index antisymmetrisation, where  
$R_{\mu\nu\ \sigma}^{\ \ \, \rho}$ is the Newton-Cartan curvature tensor 
which can easily be seen to be given by the expression
\be R_{\mu\nu\ \sigma}^{\ \ \, \rho} =2 t_\sigma t_{[\mu}\nabla_{\nu]}
g^\rho\, .\label{34a}\fe
Although it is not immediately manifest from this formula, it can be easily 
verified using (\ref{07}) and (\ref{32b}) that this curvature tensor is 
indeed invariant under the gauge transformation (\ref{06}), i.e. it satisfies 
$R_{\mu\nu\ \sigma}^{\ \ \, \rho}\mapsto \breve
R_{\mu\nu\ \sigma}^{\ \ \, \rho}$ with
\be\breve R_{\mu\nu\ \sigma}^{\ \ \, \rho}=R_{\mu\nu\ \sigma}^{\ \ \, \rho}
\, .\label{34c}\fe
It can be seen that the corresponding (again Milne gauge independent) 
Ricci type curvature trace tensor is proportional to the Laplacian of 
the gravitational potential, having only a single independent component 
that is specified by the formula 
\be R_{\mu\nu}= R_{\rho\mu\ \nu}^{\ \ \, \rho}=t_\mu t_\nu
\gamma^{\rho\sigma}\nabla_{\!\rho}\nabla_{\!\sigma}\phi
\, .\label{34b}\fe

\bigskip
{\bf 5. Action and the 4-momentum covector}
\medskip

Although, as has just been shown, the dynamical equation (\ref{33a}) 
is gauge independent, its derivation from an action principle requires 
the use of a gauge dependent momentum covector $\pi_\mu$ that is in 
many ways analogous to the electromagnetic gauge potential $A_\mu$ 
that is needed for the variational formulation of Maxwell's equation. 
The original variational formulation of the Newtonian dynamical
equation by Laplace was given by an action integral
\be {\cal I}=\int L\, dt \, ,\label{36}\fe
in which the Lagrangian scalar is taken to be the difference
between the kinetic and potential energies in the well known form
\be L={_1\over^2}m v^2-m\phi\, ,\label{36a}\fe
where $m$ is a constant mass parameter, and $v$ is the magnitude of the 
3 velocity vector $v^\mu$ as specified with respect to some chosen 
Aristotelian reference system, in terms of the corresponding ether 
vector $e^\mu$, by
\be v^\mu= u^\mu-e^\mu\, ,\hskip 1 cm v^\mu t_\mu=0\, ,\label{40}\fe
so that its Aristotelian components will be given by
\be v^{_0}=0\, ,\hskip 1 cm v^i={dX^i\over dt}\, .\label{36b}\fe
and its magnitude by 
\be v^2=\gamma_{ij}v^i v^j \, .\label{36e}\fe
This action can be rewritten in the more elegantly covariant form
\be {\cal I}=\int \pi_\mu\, dx^\mu\, ,\label{36c}\fe
which is evidently equivalent to taking
\be L=\pi_\mu u^\mu\, ,\label{36d}\fe
by defining the appropriate gauge dependent 4-momentum covector as 
follows. In terms of the Aristotelian coordinate system $x^{_0}=t$, 
$x^i=X^i$ characterizing the chosen gauge,in which we shall evidently 
have 
\be u^{_0}=1\, ,\hskip 1 cm u^i=v^i\, ,\label{37}\fe
the appropriate 4-momentum covector will be given by the prescription
\be \pi_{_0}=-{\cal E}\, ,\hskip 1 cm \pi_i= m\gamma_{ij}v^j
\, ,\label{37a}\fe
in which the total particle energy is given by the usual formula
\be {\cal E}={_1\over^2}m v^2+m\phi\, .\label{37b}\fe 
Instead of explicit reliance on the Aristotelian coordinate system, we 
can use the corresponding ether vector $e^\mu$ as introduced in 
(\ref{19b}), i.e. the unit 4-velocity vector of the Aristotelian rest
frame, with respect to which its components will be given simply by 
\be e^{_0}=1\, ,\hskip 1 cm e^i=0\, ,\label{37c}\fe
for the purpose of obtaining a covariant expression for $\pi_\mu$ as 
follows. To start with, we use the Kronecker unit tensor 
$\delta^\mu_{\ \nu}$ to construct the corresponding Aristotelian space 
projection tensor according to the specification
\be \gamma_\nu^{\,\mu}= \delta^\mu_{\ \nu} -e^{\,\mu}t_\nu
\, .\label{38}\fe
We then use the defining relations
\be \gamma_{\mu\nu} e^\nu=0\, ,\hskip 1 cm \gamma_{\mu\nu}
\gamma^{\nu\rho}=\gamma_{\mu}^{\rho} \, ,\label{38a}\fe
to characterize the covariant tensor $\gamma_{\mu\nu}$ obtained 
via the Aristotelian coordinate mapping from the space metric 
$\gamma_{ij}$ in the constant time 3-sections.
We can then define the kinetic 4-momentum vector by
\be p_\mu=m v_\mu - {_1\over ^2}m v^2 t_\mu \, ,\label{38b}\fe
where
\be v_\mu=\gamma_{\mu\nu}u^\nu\, ,\hskip 1 cm v^2=v_\mu v ^\mu
=\gamma_{\mu\nu} u^\mu u^\nu \, \label{38c}\fe
so that we have
\be {_1\over^2}m v^2= p_\nu u^\nu=-p_\nu e^\nu\, .\label{38h}\fe
Like the $u^\mu$ the kinetic momentum has only three independent
components, being subject to a constraint that in this case 
(unlike that of $u^\mu$) is ether frame dependent, having the form
\be \gamma^{\mu\nu}p_\mu p_\nu=-2me^\mu p_\mu\, .\label{38f}\fe 
In terms of this kinetic contribution we finally obtain the expression
\be \pi_\mu=p_\mu- m\phi\, t_\mu \, ,\label{38d}\fe
for the complete momentum covector.

\bigskip
{\bf 6. Finite and infinitesimal gauge transformation rules}
\medskip

Since the boost transformation law for the Aristotelian ether vector
evidently takes the form $ e^\mu\mapsto \breve e^\mu$ with
\be \breve e^\mu= e^\mu+b^\mu\, ,\label{39}\fe
it can be seen that while, the degenerate contravariant metric tensor 
is gauge invariant,
\be \breve\gamma^{\mu\nu} = \gamma^{\mu\nu}\, ,\label{39a}\fe
the corresponding mixed projection tensor (\ref{38}) will undergo a 
transformation given by
\be \breve\gamma^\mu_{\,\nu}=\gamma^\mu_{\,\nu}-t_\nu b^\mu
\, ,\label{39b}\fe
and the corresponding degenerate covariant metric tensor will be governed
by the less trivial transformation rule
\be \breve\gamma_{\mu\nu}=\gamma_{\mu\nu}-2t_{(\mu}\gamma_{\nu)\rho}
b^\rho +b^2 t_\mu t_\nu\, ,\label{39c}\fe
in which the boost magnitude $b$ is naturally defined by
\be b^2=\gamma_{\mu\nu} b^\mu b^\nu\, .\label{39d}\fe
Thus while the contravariant form of the relative velocity (\ref{40})
obeys the simple Galilean transformation rule
\be \breve v^\mu = v^\mu-b^\mu\, ,\label{40a}\fe
the corresponding covariant vector
\be v_\mu=\gamma_{\mu\nu}v^\nu \label{40b}\fe 
transforms according to the less simple rule
\be \breve v_\mu= v_\mu -\gamma_{\mu\nu} b^\nu +t_\mu
(b^2-b^\nu v_\nu)  \, ,\label{40c}\fe
while for the squared velocity 
\be v^2=v^\mu v_\mu= \gamma_{\mu\nu} u^\mu u^\nu\, ,\label{40d}\fe
one obtains
\be \breve v^2 = v^2-2 b_\mu v^\mu +b^2\, .\label{40e}\fe

When applied to the kinetic momentum covector (\ref{38b}) the foregoing
formulae provide the gauge transformation rule
$p_\mu\mapsto \breve p_\mu$ with
\be \breve p_\mu = p_\mu-m\gamma_{\mu\nu}b^\nu+{_1\over^2}m b^2 t_\mu
\, ,\label{41}\fe
and thus the transformation rule for the complete momentum covector 
(\ref{38d}) can be seen to be expressible in terms of the boost 
potential $\beta$ by
\be \breve \pi_\mu = \pi_\mu-m\nabla_{\!\mu}\beta +{_1\over^2}
m \, b^2 t_\mu\, ,\label{41a}\fe
so that, in particular, the corresponding transformation law for its 
energy component (\ref{37b}) will be expressible as
\be \breve{\cal E} = {\cal E}-p_\nu b^\nu +m\big({_1\over^2}b^2+\alpha
\big)\, ,\label{41b}\fe
while finally for the Lagrangian scalar (\ref{36d}) we obtain
$L\mapsto \breve L$ with
\be \breve L = L-mu^\mu\nabla_{\!\mu}\beta+{_1\over^2}m\, 
b^2\, .\label{41c}\fe

It is apparent from the form of these transformation rules that it will be 
convenient to work with a recalibrated boost potential,
$\hat\beta=\beta-{1\over2}\int b^2 dt$, that will be characterized by
\be \nabla_ {\!\nu}\hat\beta=\nabla_{\!\nu}\beta-{_1\over^2}
b^2 t_\nu\, .\label{18}\fe
Since the difference between $\beta$ and $\hat\beta$ is a function only
of the cosmological time, we can just as well use the latter as the former 
in the characterization (\ref{19}), which can be rewritten as
\be b^\mu=\gamma^{\mu\nu}\nabla_{\!\nu}\hat\beta \, ,\hskip 1 cm
\gamma^{\mu\nu}\nabla_{\!\nu} b^\rho=0 \, ,\label{18a}\fe
but in terms of the recalibrated boost potential the expression (\ref{19a}) 
for the acceleration will acquire the slightly modified form
\be a^\mu=\gamma^{\mu\nu}\nabla_{\!\nu}\alpha \, ,
\hskip 1 cm \alpha= e^\mu\nabla_{\!\mu}\hat\beta + {_1\over^2}
b^2\, .\label{18b}\fe
The corresponding modification of the formula (\ref{41b}) for the
transformation of the energy will be given by
\be \breve{\cal E}={\cal E}-p_\nu b^\nu +m(e^\nu+b^\nu)\nabla_{\!\nu}
\hat\beta\, .\label{18c}\fe
However it is for the 4-momentum covector and the Lagrangian that the 
advantage of the modified boost potential becomes apparent, since it allows 
(\ref{41a}) and (\ref{41b}) to be rewritten more simply as
\be\breve \pi_\mu = \pi_\mu- m\nabla_{\!\mu}\hat\beta\, ,\label{18d}\fe
and
\be \breve L = L- mu^\mu\nabla_{\!\mu}\hat\beta\, .\label{18e}\fe

The effect of the gauge transformation on the action integral (\ref{36}) 
can thus be seen to be given by ${\cal I}\mapsto \breve{\cal I}$ with
\be \breve{\cal I} = {\cal I}-m\Big[\hat\beta\Big] \, ,\label{41d}\fe
using large square brackets to denote the total change in the boost 
potential $\beta$ along the worldline segment under consideration. Since
the gauge adjustment term in (\ref{41d}) will not be affected by any purely 
local variation of the worldline (local meaning that it is non vanishing
only on a confined subsegment lying entirely inside the extended worldline 
segment under consideration) it is obvious that it will have no effect as 
far as the application of the variational principle is concerned. The 
observation that the gauge transformation changes the action only by an 
amount that is constant in the sense of being independent of local worldline 
variations evidently accounts for the invariance with respect to these 
(linear Galilean or accelerated Milne) transformations of the ensuing system 
of dynamical equations.

The preceeding example is one of many in which it is simpler not to work 
with finite (Galilean or Milne) transformations as we have been doing so 
far, but with linearized infinitesimal transformations. For a given gauge 
transformation generated by a given boost potential $\beta$, whereby a 
generic quantity, $q$ say, is subject to a mapping $q\mapsto\breve q$, the 
corresponding infinitesimal gauge transformation $q\mapsto \breve{\rm  d}q$ 
is defined by a routine two step procedure as follows. The first step is to
construct a homotopic interpolation by a one parameter family of gauge 
transformations $q\mapsto q\{\epsilon\}$ with $q\{0\}=q$ and 
$q\{1\}=\breve q$, for which $q\{\epsilon\}$ is given, for intermediate 
values of the homotopy parameter $\epsilon$, by an interpolating boost 
potential $\beta\{\epsilon\}=\epsilon\beta$. The corresponding 
infinitesimal transformation is then obtained by taking the limit, as 
$\epsilon\rightarrow 0$, of the derivative with respect to $\epsilon$, so 
that we have
\be  \breve {\rm d} q=\lim_{\epsilon\rightarrow 0}
{{\rm d}\over {\rm d}\epsilon}\Big( q\{\epsilon\}\Big) \, .\label{96f}\fe

This differential operation will merely restore the value we started with 
(as recovered by setting $\epsilon=0$) for quantities whose dependence on 
the transformation amplitude is homogeneously linear as is the case for the 
diverse derivatives of the boost function, which will be characterized by
\be \breve {\rm d}\alpha=\alpha\, ,\hskip 1 cm
\breve {\rm d} b^\mu=b^\mu\, ,\hskip 1 cm \breve{\rm d }
 a^\mu=a^\mu\, ,\label{96}\fe
and also of course for the original boost function $\beta$ itself,
though not for the modified boost function $\hat\beta$ for which
we obtain a convenient simplification,
\be  \breve{\rm d} \hat\beta= \breve {\rm d} \beta  =\beta\, ,\label{96a}\fe
which means that at the differential level the modification $\hat\beta$ is 
redundant. In a similar manner, going to the differential level provides 
no great simplification for quantities whose gauge transformation depends 
just linearly on the boost amplitude, as is the case for ether vector
$e^\mu$, and the 3-velocity vector $v^\mu$, for which we obtain
\be \breve {\rm d} e^\mu=b^\mu\, ,\hskip 1 cm
\breve {\rm d} v^\mu =-b^\mu\, ,\label{96b}\fe
as well as for the less trivial cases of the mixed projection tensor 
$\gamma^\mu_\nu$, and the connection $\Gamma_{\!\mu\ \rho}^{\ \nu}$,
for which, from (\ref{39b}) and (\ref{06}), we obtain
\be \breve {\rm d} \gamma^\mu_\nu=-t_\nu b^\mu\, ,\hskip 1 cm
\breve {\rm d}\Gamma_{\!\mu\ \rho}^{\ \nu}= -t_\mu a^\nu t_\rho
\, ,\label{06a}\fe
but for quantities with more complicated gauge dependence the differential
level is much more convenient. Noteworthy examples are the covariant space 
metric tensor $\gamma_{\mu\nu}$ and the kinetic momentum vector $p_\mu$ 
for which the formulae (\ref{39c}) and (\ref{41}) reduce simply to
\be \breve {\rm d}\gamma_{\mu\nu}=-2t_{(\mu}\gamma_{\nu)\rho}
b^\rho \, ,\hskip 1 cm  \breve {\rm d} p_\mu = -m\gamma_{\mu\nu}b^\nu
\, ,\label{96c}\fe
and particularly the Lagrangian $L$, and the complete momentum covector 
$\pi_\mu$, for which we simply obtain
\be \breve {\rm d} L=-m u^\nu\nabla_{\!\nu}\beta\, ,\hskip 1 cm
 \breve {\rm d}\pi_\mu=- m\nabla_\mu\beta\, .\label{96d}\fe
(The feature of transforming  just by the addition of the gradient of a 
scalar is a property that the 4-momentum covector $\pi_\mu$ of Newtonian
dynamics shares with the covector potential $A_\mu$ of Maxwellian
electromagnetism, a relationship that is essential for the amalgamation
of gravity and electromagnetism in the corresponding general
relativistic theory~\cite{Carter89}.) 

The corresponding differential version of the formula (\ref{41d}) for the 
action integral is given by the expression
\be \breve {\rm d}{\cal I} = -m\Big[\beta\Big] \, ,\label{96e}\fe
whose evident path independence makes it obvious that the ensuing dynamical 
theory will have to be gauge invariant, as shown above by the existence of 
the manifestly covariant formulation (\ref{33a}) of the equations of motion.

\vfill\eject
\bigskip
{\bf 7. Covariant fluid current variation formulae.}
\medskip

The simple Lagrangian worldline variation principle for a single particle
that was discussed in the preceeding sections can be generalised in an 
obviously natural way to a flow line variation principle for a fluid 
system.

Before proceeding, it is to be recalled that in the absence of any
prescribed spacetime structure the usual description of currents in terms
of vector fields will not be available, but it will still be possible
to use the more fundamental Cartan type description whereby a current
is represented as a 3-form, i.e. an antisymmetric covariant tensor
with 4-dimensional components $N_{\mu\nu\rho}$ whose surface
integral
\be N={1\over 3!}\int N_{\mu\nu\rho}\, d^3 x^{\mu\nu\rho}
\label{43}\fe
determines the total number flux over a 3-surface with tangent element
 $d^3x^{\mu\nu\rho}$ $= 3! d_{_{(1)}} x^{[\mu}\, d_{_{(2)}} x^\nu
d_{_{(3)}}x^{\rho]}$ generated by infinitesimal displacements
$d_{_{(i)}} x^\mu$ ( $i$=1,2,3). In such a description, the condition
for the conservation of the number flux is the vanishing of its 
exterior derivative as defined by $(\partial\wedge N)_{\mu\nu\rho\sigma}$
$=4\partial_{[\mu}N_{\nu\rho\sigma]}$. A convenient feature of
such exterior differentiation is that it makes no difference if the
partial differentiation operator $\partial_\mu$ is replaced by a 
tensorially covariant differentiation operator $\nabla_{\!\mu}$
(or by Cartan's gauge covariant differentiation operator $D_\mu$) since due 
to the antisymmetrisation all the (symmetric) connection components 
will cancel out.

The more fundamental three index description of the current will of 
course be replacable by a more compact description involving a single 
contravariant index whenever the spacetime background is endowed with a 
canonical antisymmetric measure tensor $\varepsilon_{\mu\nu\rho\sigma}$ 
and a corresponding contravariant alternating tensor 
$\varepsilon^{\mu\nu\rho\sigma}$, which, as seen above, will be the 
case both in relativistic and Newtonian spacetime. The required current 
vector, with Aristotelian rest frame components
\be n^{_{0}}=n\, ,\hskip 1 cm n^i= n v^i\, ,\label{43a}\fe
where $n$ is the ordinary (scalar) particle number density, will then 
be given by the duality relation
\be n^\mu={1\over 3!}\varepsilon^{\mu\nu\rho\sigma} 
N_{\nu\rho\sigma}\, ,\label{43b}\fe
which can be inverted to provide the expression
\be N_{\mu\nu\rho}=\varepsilon_{\mu\nu\rho\sigma} n^\sigma
\, .\label{43c}\fe
For {\it any} covariant symmetric connection compatible with the measure
preservation condition (\ref{20a}), the corresponding covariant 
differentiation operator will determine a divergence that will just be 
the dual of the exterior derivative operator. Thus independently of the 
choice of gauge one obtains the equivalent expressions
\be \nabla_{\!\mu} n^\mu=D_\mu n^\mu={1\over 4!}
\varepsilon^{\mu\nu\rho\sigma} (\partial\wedge N)_{\mu\nu\rho\sigma}=
{1\over 3!}\varepsilon^{\mu\nu\rho\sigma}N_{\nu\rho\sigma ,\mu}
\, ,\label{44}\fe
for the particle rate, which will vanish for a current that is conserved.

In order to apply the variational principle, we need to evaluate the
variation of the current that will result from transport by the
action of an infinitesimal displacement $x^\mu\mapsto x^\mu+ \xi^\mu$.
As a general principle~\cite{Carter89} the fixed point (Eulerian) variation
of any field will be given generally by its comoving (Lagrangian)
variation minus the relevant Lie derivative. Since the specification
of the covariant representation of the current does not depend on any
background structure its comoving variation will simply vanish, so its
fixed point variation will just be given by the formula
\be \delta N_{\mu\nu\rho}=-\vec \xi \Libra N_{\mu\nu\rho}
\, ,\label{45}\fe
in which the Lie derivative is given by the standard formula
\be \vec \xi \Libra N_{\mu\nu\rho} = \xi^\sigma\nabla_{\!\sigma}
N_{\mu\nu\rho}+3N_{\sigma[\mu\nu}\nabla_{\!\rho]}\xi^\sigma
\, .\label{45a}\fe

As for any Lie derivative formula (and as in exterior differentiation)
it makes no difference if the partial differentiation operator
$\partial_\mu$ is replaced by a tensorially covariant differentiation
operator $\nabla_{\!\mu}$ (or by Cartan's gauge covariant differentiation
$D_\mu$) since due to the antisymmetrisation all the (symmetric)
connection components will cancel out. Thus by taking the dual of the
formula (\ref{45a}) we obtain the useful theorem that for {\it any}
covariant symmetric connection compatible with the measure preservation
condition (\ref{20a}), the corresponding covariant differentiation
operator $\nabla_{\!\mu}$ can be used to express the fixed point
(Eulerian) current variation produced by the displacement vector field
$\xi^\mu$ in the form
\be \delta n^\mu= n^\nu\nabla_{\!\nu} \xi^\mu-\xi^\nu\nabla_{\!\nu}
n^\mu - n^\mu\nabla_{\!\nu} \xi^\nu\, .\label{45b}\fe
This result establishes the validity in a Newtonian framework of a
formula that has long been in regular use in a relativistic context,
where it was originally derived by a rather different line of
reasonning~\cite{Carter89} that depended on the Riemannian specification
of the covariant differentiation in terms of the non-degenerate space
time metric that is no longer available in the Newtonian case.
It is clear from this present approach that the formula (\ref{45b})
will remain valid if Cartan's gauge covariant derivative operator
$D_{\!\mu}$ is substituted in place of the flat but gauge dependent
derivative operator $\nabla_{\!\mu}$.

A useful corollary of (\ref{45b}) is the corresponding formula for the
variation of the current divergence, which takes the simple form
\be \delta\big(\nabla_{\!\nu}n^\nu\big)=-\nabla_{\!\mu}\big(
\xi^\mu\nabla_{\!\nu}n^\nu\big)\, .\label{45c}\fe
It is immediately apparent from this that if the original current
is conserved, i.e. if $\nabla_{\!\nu}n^\nu$ vanishes, then the displaced
current will have the same conservation property.

The formal identity of the relativistic and Newtonian variation
formulae will be lost if we make a decomposition of the usual form
\be n^\mu=n u^\mu\, ,\hskip 1 cm n=n^\mu t_\mu \, ,\label{46}\fe
in which $n$ is the ordinary particle number density scalar and
$u^\mu$ is the 4-velocity of the flow as characterized by the
unit normalisation condition (\ref{31a}). By contracting (\ref{45b})
with $t_\mu$ it can seen that the variation law for the particle number
density will be given by
\be \delta n=t_\mu n^\nu\nabla_{\!\nu}\xi^\mu-\nabla_{\!\nu}(n\xi^\nu)
\, ,\label{46a}\fe
which has a different form from its relativistic analogue
~\cite{Carter89} due to its dependence on the preferred time basis
vector $t_\mu$ (instead of the non degenerate spacetime metric
$g_{\mu\nu}$ that plays the corresponding role in the relativistic
version).
The same remark applies to the corresponding variation of
the 4-velocity of the flow, which can be seen from (\ref{45b}) and
(\ref{46a}) to be given by
\be \delta u^\mu= u^\nu\nabla_{\!\nu} \xi^\mu -\xi^\nu\nabla_{\!\nu}
u^\mu -u^\mu u^\nu t_\rho\nabla_{\!\nu}\xi^\rho\, .\label{46b}\fe

\bigskip
{\bf 8. Action principle for simple perfect fluid.}
\medskip

The natural way to extend the single particle action principle discussed
above to a corresponding fluid action principle is to base the latter on
the space time integral of a Lagrangian density $\Lambda$ that will be
given by a decomposition of the form
\be \Lambda=\Lambda_{_{\rm pot}}+\Lambda_{_{\rm kin}}
+\Lambda_{_{\rm int}} \, ,\label{47} \fe
in which the first two (gauge dependent) terms are given by the product
of the relevant particle number density $n$ with the corresponding single
particle contributions, while the extra
term $\Lambda_{_{\rm int}}$ is given simply by
\be \Lambda_{_{\rm int}}= -U_{_{\rm int}}\, ,\label{47a} \fe
where $U_{_{\rm int}}$ is the internal
compression energy, which will be a (naturally gauge independent) function
just of the particle number density $n$. Specifically, the external
potential energy contribution will be given by usual Newtonian formula
\be \Lambda_{_{\rm pot}}= -nm\phi  \, ,\label{48}\fe
which may be rewritten in covariant form as
\be \Lambda_{_{\rm pot}}=-\phi\, \rho^\mu t_\mu\, ,\label{48a}\fe
where the mass density current is defined by
\be \rho^\mu=\rho u^\mu = m n^\mu\, ,\hskip 1 cm \rho=nm\, ,\label{48b}\fe
while in accordance with (\ref{38h}) the kinetic contribution will be given by
\be \Lambda_{_{\rm kin}}= n p_\mu u^\mu = n^\mu p_\mu\, ,\label{48c}\fe
in terms of the (gauge dependent) kinetic 4-momentum covector defined
by (\ref{38b}). The extra (gauge independent) contribution (\ref{47a})
representing the negative of the internal compression  energy density will
determine a corresponding (gauge independent) chemical potential $\chi$
by a variation formula of the standard form
\be \delta U_{_{\rm int}}= \chi\delta n \, .\label{49}\fe
In terms of this chemical potential function, the relevant perfect fluid
pressure function $P$ (which is also gauge independent) will be given by
the well known formula
\be P=n\chi-U_{_{\rm int}}\, .\label{60}\fe

Since, as a consequence of the restriction (\ref{38f}), the variation of the
kinetic momentum is automatically constrained to satisfy the identity
\be u^\mu\delta p_\mu=0 \, ,\label{49a}\fe
it follows that the generic variation of the purely kinetic contribution
to the Lagrangian will be given simply by
\be \delta \Lambda_{_{\rm kin}}=p_\mu\delta n^\mu\, ,\fe
The variation of the combination  (\ref{47}) will therefore be given by
an expression of the canonical form
\be \delta \Lambda=\pi_\mu \delta n^\mu -\rho\delta\phi \, ,\label{49b}\fe
in which the total 4-momentum is given by an expression of the form
\be \pi_\mu=\mu_\mu - m\phi\, t_\mu\, ,\label{50}\fe
which differs from the corresponding free particle momentum formula
(\ref{38d}) by the replacement of the purely kinetic contribution $p_\mu$
by a total material 4-momentum covector $\mu_\mu$ that is defined by
\be \mu_\mu =p_\mu -\chi t_\mu\, .\label{50a}\fe
This material momentum covector is alternatively definable by the
variation formula
\be\delta\Lambda_{_{\rm mat}}=\mu_\mu\delta n^\mu\, ,\label{50b}\fe
where the total material (non gravitational) contribution to the
Lagrangian density is defined by
\be \Lambda_{_{\rm mat}}=\Lambda_{_{\rm kin}}+\Lambda_{_{\rm int}}
\, .\label{50c}\fe

It can now be seen that the complete Lagrangian (\ref{47}) will be elegantly
expressible in terms of the 4-momentum covector $\pi_\mu$ and the pressure
function $P$ as
\be \Lambda= n^\mu\pi_\mu+ P\, .\label{50d}\fe
It is to be remarked that while the pressure term in (\ref{50d}) is gauge
invariant, the first term is not. However as the formula (\ref{50}) for the
fluid particle 4-momentum covector $\pi_\mu$ differs from its single particle
analogue (\ref{38d}) only by the gauge independent term proportional to
$\chi$ in (\ref{50a}), it can be seen its variation $\breve {\rm d} \pi_\mu$
under the action of an infinitesimal gauge transformation will be given by
the same simple formula (\ref{96d}) as in the single particle case. It
follows that the corresponding infinitesimal gauge variation of the
Lagrangian density (\ref{50d}) will be given simply by
\be \breve {\rm d}\Lambda=-\rho^\nu\nabla_{\!\nu}\beta\, ,\label{50e}\fe
If it is taken for granted that the fluid obeys the ordinary Newtonian
mass conservation law
\be \nabla_{\!\nu}\rho^\nu= 0\, ,\label{63b}\fe
the infinitesimal gauge variation will be expressible as a pure divergence
in the form
\be \breve {\rm d}\Lambda=-\nabla_{\!\nu}\big(\beta\rho^\nu\big)
\, .\label{63c}\fe
This means that the gauge change will have no effect on a localised
variation of the space-time volume integral of the action density, so the
dynamical equations given by the action principle will automatically be
gauge independent.

For the actual evaluation of the variation of the action, the work of the
preceeding section provides all the elements that are needed. It can be
seen from (\ref{49b}) and (\ref{45b}) that when the fluid flow is subjected
to the action of an infinitesimal displacement vector field $\xi^\mu$, the
resulting variation of the Lagrangian density will be given by the formula
\be \delta\Lambda =\nabla_{\!\mu}\big(2\pi_\nu n^{[\mu}\xi^{\nu]}
\big) -f_\mu \xi^\mu -\rho\delta\phi\, ,\label{51}\fe
in which the covector $f_\mu$ is interpretable as the 4-force density
acting on the fluid, excluding the gravitational contribution which is
already taken into account within the formalism. This force density can be
seen to be given by the prescription
\be f_\mu=2n^\nu \nabla_{\![\nu}\pi_{\mu]}+\pi_\mu\nabla_{\!\nu} n^\nu
\, ,\label{51a}\fe
i.e. it is constructed from the current vector and the corresponding
momentum covector by contraction with (exterior) derivative plus derivative
of contraction.

The postulate that the mass parameter $m$ should be constant means that
the mass conservation law (\ref{63b}) will be equivalent to the particle
conservation law given, in accordance with (\ref{44}) by
\be \nabla_{\!\nu} n^\nu=0\, ,\label{59a}\fe
a result that is alternatively derivable, wherever the product
$n^\mu\pi_\mu$ is non zero,  as a consequence of the variational requirement
that the force density (\ref{51a}) should vanish. Subject to (\ref{59a}),
the expression for the force density will evidently reduce to the simple
form
\be f_\mu=n^\nu {\varpi}_{\nu\mu} \, ,\label{52}\fe
in which the relevant generalized 4-vorticity 2-form is defined, as the
exterior derivative of the 4-momentum covector, by
\be \varpi_{\mu\nu}=2\nabla_{\![\mu}\pi_{\nu]}\, .\label{51b}\fe
This vorticity 2-form $\varpi_{\mu\nu}$ is generalized in the sense that it
automatically includes allowance for gravity, whose effect can be separated
out in the decomposition
\be \varpi_{\mu\nu}=w_{\mu\nu}+2m t_{[\mu}\nabla_{\!\nu]}\phi
\, .\label{52a}\fe
in which $w_{\mu\nu}$ is the ordinary material vorticity tensor defined by
\be w_{\mu\nu}=2\nabla_{\![\mu}\mu_{\nu]}\, .\label{52b}\fe
The adjustment allowing for gravitation affects only the time components,
so both the complete and the material vorticity give the same purely
spacelike 3-velocity vector, which is expressible independently of $\phi$ as
\be w^\mu={_1\over^2}\varepsilon^{\mu\nu\rho}w_{\nu\rho}=
{_1\over^2}\varepsilon^{\mu\nu\rho}\varpi_{\nu\rho} \, ,\label{58b}\fe
and which is related to the purely kinematic local angular velocity
vector $\omega^\mu$ by the proportionality relation
\be w^\mu=2m\omega^\mu\, ,\hskip 1 cm \omega^\mu={_1\over^2}
\varepsilon^{\mu\nu\rho}\nabla_{\!\nu}v_\rho\, .\label{58c}\fe

A related quantity that is easy to analyse in the covariant formalism
we are using here, but much more awkward to treat using the traditional
3+1 spacetime decomposition -- so much so that its role in Newtonian fluid
dynamics, was not recognised until the relatively recent work of
Moreau~\cite{Moreau61} and Moffat~\cite{Moffat69} (a century after the
pionnering analysis of vorticity by workers of Kelvin's generation) is
that of helicity. In that work (and in its more recent non-barotropic
generalisation~\cite{Gaffet85}) the helicity was introduced as a scalar
density that was constructed as the three dimension scalar product of the
velocity vector $v^i$ and the vorticity vector $w^i$. On the basis of
experience~\cite{Carter89} with the relativistic case, it is evident that in
the 4-dimensionally covariant formalism we are using here, the helicity will
most naturally be definable ~\cite{Peradzynski90,CarterKhalatnikov94} as a
vectorial current $\eta^\mu$ that is proportional to the dual of the
exterior product of the energy momentum covector $\pi_\mu$ with the
corresponding generalized vorticity two form, $\varpi_{\mu\nu}$, namely
\be \eta^\mu=\varepsilon^{\mu\nu\rho\sigma}\pi_\nu\nabla_{\!\rho}
\pi_\sigma={_1\over^2}\varepsilon^{\mu\nu\rho\sigma}\pi_\nu
 \varpi_{\rho\sigma}\, .\label{58d}\fe
The time component of this quantity can be seen from (\ref{04}) to
be proportional to the Moreau-Moffat helicity scalar, with a negative
coefficient (in the sign convention we are using, for which the sign of
the measure component $\varepsilon_{_{0123}}$ is taken to be positive,
so that of $\varepsilon^{_{0123}}$ will be negative) that is given by
$-2m^2$, i.e.
\be \eta^\mu t_\mu =-w^\mu\pi_\mu =-2 m^2\omega^i v_i
\, .\label{58e}\fe
It immediately follows from the Eulerian dynamical equation, whose
variational derivation will be described below, that this helicity
current $\eta^\mu$ will be conserved in the simple sense that its
4-divergence,
\be\nabla_{\!\mu}\eta^\mu={_1\over^4}\varepsilon^{\mu\nu\rho\sigma}
\varpi_{\mu\nu}\varpi_{\rho\sigma}\, ,\label{58f}\fe
will turn out to vanish, a property that is both obvious and easy
to express in the 4-dimensional approach used here, but not so trivial,
either to derive or even to present, within the (Latin as opposed to
Greek index) framework of the traditional 3+1 formalism (see appendix).

The purport of the variational principle is that the spacetime volume
integral of $\delta\Lambda$ should vanish for any displacement $\xi^\mu$
with bounded support (i.e. that vanishes outside some bounded spacetime
region). Since, by Green's theorem, the divergence contribution in
(\ref{51}) will make no contribution to the variational integral, the
principle reduces to the requirement that the 4-force density should vanish,
\be f_\mu=0\, .\label{59}\fe
Subject to (\ref{59a}) this equation will reduce to the form
\be n^\nu {\varpi}_{\nu\mu}=0 \, ,\label{59b}\fe
in which the complete vorticity 2-form can be expressed as
\be \varpi_{\mu\nu}=2\nabla_{\![\mu}p_{\nu]}
+2 t_{[\mu}\nabla_{\!\nu]}(\chi+m\phi)\, .\label{59c}\fe
An immediate consequence of (\ref{59b}) is that (since its components form
an antisymmetric matrix with a zero eigenvalue eigenvector, namely $n^\mu$)
the vorticity 2-form $\varpi_{\mu\nu}$ must be algebraicly degenerate, with
vanishing determinant and therefore matrix rank 2 (since an antisymmetric
matrix cannot have even rank), a condition that is expressible by the
algebraic restriction
\be \varpi_{\mu[\nu}\varpi_{\rho\sigma]}=0 \, .\label{58g}\fe
This has the obvious corollary that the right hand side of (\ref{58f}) must
vanish, and hence that the helicity current $\eta^\mu$ will indeed be
conserved, i.e. we shall have
\be\nabla_{\!\mu}\eta^\mu=0\, .\label{59f}\fe
Another, more widely known, consequence of the degeneracy property
(\ref{58g}) is that (exactly as in the relativistic case explained
elsewhere~\cite{Carter89}) the vorticity 2-form will be orthogonal to a 2
dimensional tangent element, containing the flow vector $n^\mu$ as well as
the vorticity 3-vector $w^\mu$ defined by (\ref{58b}), and that (as in the
analogous case of magnetic 2-surfaces in a perfectly conducting plasma)
these 2-dimensional tangent elements will be integrable in the sense of
meshing together to form well defined vorticity flux 2-surfaces.

The analogue for a 2-form $\varpi_{\mu\nu}$ of the formula
(\ref{45a}) for Lie derivation with respect to a flow field
$n^\mu$ takes the form
\be \vec n\Libra {\varpi}_{\mu \nu} = 3 n^{\sigma}
\nabla_{\![\sigma}\varpi_{\mu\nu]}-2\nabla_{\![\mu}(\varpi_{\nu]\sigma}
n^\sigma)\, ,\label{59k}\fe
in which the first term will drop out identically by the closure
property, i.e. the vanishing of the exterior derivative
$3\nabla_{\![\sigma}\varpi_{\mu\nu]}$ of the vorticity as an
automatic consequence of its exactness property (\ref{51b}). When
the field equation (\ref{59b}) is satisfied, it can be seen
that the second term in (\ref{59k}) will also drop out. We thus
obtain the covariant generalisation of the well known Kelvin
vorticity conservation theorem ~\cite{Landau} to the effect
that the the vorticity 2-form will be conserved by Lie transport,
with respect to any arbitrarily rescaled multiple of the
flow vector, i.e. 
\be (\zeta \vec n)\Libra {\varpi}_{\mu \nu} = 0\, ,\label{59l}\fe 
for an arbitrarily variable scalar field $\zeta$.

It is to be noted that if, instead of restricting the variation
$\delta n^\nu$ to be given by the worldline displacement formula
(\ref{45b}), one merely imposes current conservation by adding
a Lagrange multiplier term $\varphi\nabla_{\!\mu} n^\mu$ to the
action density, then one will get a more restricted dynamical
equation to the effect that  the momentum covector should be the
gradient of the Lagrange multiplier $\varphi$ and thus that it
should be irrotational:
\be \pi_\mu=\nabla_{\!\mu}\varphi\hskip 1 cm \Rightarrow\hskip 1 cm
\varpi_{\mu\nu}=0\, .\label{59i}\fe
A solution of this irrotational type is the only kind that is allowed
in the special case of a simple superfluid (on a mesoscopic -- i.e.
intervortex -- scale) for which the scalar $\varphi$ is interpretable
as being proportional to the quantum phase angle of a bosonic condensate.

In terms of the Newton-Cartan connection given by (\ref{33}) the dynamical
equation (\ref{59b}) can be rewritten in the manifestly gauge invariant form
\be u^\nu D_{\!\nu}u^\mu=-{1\over m}\gamma^{\mu\nu}\nabla_\nu\chi
\, .\label{59d}\fe
This is just a covariant reformulation of the well known Euler equation,
which is traditionally expressed in terms of the pressure function
(\ref{60}), whose variation can be seen from (\ref{49}) to be given in
terms of that of the chemical potential $\chi$ by
\be\delta P=n\delta\chi \, .\label{60a}\fe
It is thereby possible to rewrite (\ref{59d}) as
\be u^\nu\nabla_\nu u^\mu=-\gamma^{\mu\nu}\Big(\nabla_\nu\phi
+{1\over\rho}\nabla_\nu P\Big)\, ,\label{60b}\fe
which can immediately be translated into Aristotelian coordinate notation
to give the original Eulerian version in the form~\cite{Landau}
\be \nabla_{_0}v_i +v^j\nabla_j v_i= -\nabla_{\!i}\phi-{1\over\rho}
\nabla_{\!i}P\, .\label{60c}\fe

       While this last version has the advantage of familiarity, and
(\ref{59d}) has the advantage of manifest gauge covariance (with respect
to linear Galilean and non-linear Milne transformations) it is the version 
(\ref{59b}) that is most convenient for many mathematical purposes, since 
it involves only exterior differentiation, and can therefore be evaluated 
in arbitrarily curved (e.g. comoving) coordinates using only partial 
differentiation, without reference to any of the various relevant 
connections (the  $\omega_{\mu\ \rho}^{\ \, \nu}$ that is covariant but 
curved or the connection $\Gamma_{\!\mu\ \rho}^{\ \nu}$ that is flat but 
gauge dependent). An example is the demonstration above of the
way the use of (\ref{59b}) greatly facilitates the treatment of helicity  
conservation, a concept that is almost trivial (actually simpler than the
concept of vorticity conservation) in the covariant formalism developed 
here, but  whose original derivation, in the frame dependent notation of 
(\ref{60c}) was of a technical complexity such that it was finally 
obtained (by Moreau and Moffat in the 1960's~\cite{Moreau61,Moffat69}) 
about a century later than the development of the more elementary precursor 
concept of vorticity (by nineteenth century pionneers such as Kelvin). The 
advantage of the fully covariant approach will be even greater when we go on 
from simple to multiconstituent fluids.

\bigskip
{\bf 9. Multiconstituent fluid models}
\medskip

We now extend the discussion to cases involving several independent
-- but not always independently conserved -- currents with current 
4-vectors that we shall denote by
$n_{_{\rm X}}^{\, \nu}$ where $_{\rm X}$ is a ``chemical'' index
with values ranging over the labels of the various constituents involved. 
In particular the neutron star application for which this work is 
particularly intended, will involve a neutron number density current 
$n_{_{\rm n}}^{\, \nu}$ and a proton number density current 
$n_{_{\rm p}}^{\, \nu}$ so in this case the index $_{\rm X}$ will range 
over the pair of values $_{\rm X}= {\rm n}$ and $_{\rm X}={\rm p}$. 
Although the total baryon number current $n_{_{\rm b}}$ 
$=n_{_{\rm n}}+n_{_{\rm p}}$ will be conserved, in applications 
dealing with long term evolution the neutron and proton currents 
will not be separately conserved due to the possibility of transfer 
of baryons from one to the other by weak interactions. To deal with 
such cases it may be necessary to allow for the possibility that a 
particular current $n_{_{\rm X}}^{\,\nu}$ may be characterized by a non 
vanishing value of the destruction rate (per unit spacetime volume) that 
is defined (as the negative of the corresponding creation rate) by
\be {\cal D}_{_{\rm X}}=-\nabla_{\!\nu} n_{_{\rm X}}^{\, \nu}
\, ,\label{61}\fe
a formula in which, as explained above, it makes no difference what frame 
may have been used to specify the connection involved in the covariant 
differentiation operator $\nabla_{\!\nu}$.
 
The obviously natural way to set up an appropriate Lagrangian for a 
multiconstituent fluid model is to take a combination of the same general 
form 
\be \Lambda=\Lambda_{_{\rm mat}}+\Lambda_{_{\rm pot}}\, ,\label{62}\fe 
as before, with the material contribution $\Lambda_{_{\rm mat}}$ again 
given by a decomposition of the form (\ref{50c}) as a sum of kinetic 
and internal contributions. As always, the gravitational potential energy 
contribution will simply be given by
\be \Lambda_{_{\rm pot}} =-\phi\, \rho \, ,\label{62a}\fe
where $\rho$ is the total mass density. However this will now be given as 
a sum over constituents of the form
\be \rho = {_\sum \atop ^{_{\rm X}}} m^{_{\rm X}}n_{_{\rm X}}\, ,\label{63}\fe
in which $m^{_{\rm X}}$ is the Newtonian mass per particle associated 
with the current $n_{_{\rm X}}^{\,\nu}$. The prescription (\ref{62a}) 
evidently has the same general form (\ref{48a}) as before when written 
covariantly in terms of the corresponding total mass current vector
\be \rho^\nu = {_\sum \atop ^{_{\rm X}}} m^{_{\rm X}}n_{_{\rm X}}^\nu 
\, .\label{63a}\fe
One of the basic principles of Newtonian theory is that although the 
different contributions need not be separately conserved (as matter can 
be transferred from one to another by chemical or nuclear reactions) the 
total mass current (including all relevant contributions) will still have 
to obey the conservation law (\ref{63b}).

In summation formulae such as (\ref{63}) and (\ref{63a})  (in which it 
would be legitimate to use the standard shorthand summation convention 
whereby the explicit use of the summation symbol $\Sigma$ is omitted) it is 
to be noticed that the constituent indices of the masses have been written 
``upstairs'' to indicate their contravariant character with respect to 
linear constituent recombinations, in contrast with the currents, with 
indices ``downstairs'', which undergo recombinations of the corresponding 
covariant (inverse) form. The formulae (\ref{63}) and (\ref{63a}) can be 
seen to be covariant, while the resulting sums $\rho$ and $\rho^\mu$ 
themselves are actually invariant, when such linear  transformations of 
chemical basis are carried out.  

A simple illustration of a change of chemical basis is provided by typical 
astrophysical applications for which it may be sufficient to treat the 
relevant matter (e.g. in a stellar atmosphere) as a mixture of hydrogen 
(with atomic nucleus containing just one proton) and helium (with atomic 
nucleus consisting of 2 protons and 2 neutrons), so that in terms of 
chemical index values $_{\rm X}= _{\ \rm H}$ and $_{\rm X}=_{\ \rm He}$ 
the total mass density will be given by
$\rho=m^{_{\rm H}}n_{_{\rm H}}+m^{_{\rm He}}n_{_{\rm He}}$.
In an equivalent description based on the underlying proton and neutron 
number densities, using index values $_{\rm X}= {\,\rm p}$ and 
$_{\rm X}={\,\rm n}$, the total mass density will be given by an 
expression of the exactly analogous form
$\rho=m^{_{\rm p}}n_{_{\rm p}}+m^{_{\rm n}}n_{_{\rm n}}$
in which the relevant densities are provided by a chemical basis 
transformation that is given by the relations $n_{_{\rm p}}$
$=n_{_{\rm H}}+2n_{_{\rm He}}$  and  $n_{_{\rm n}}$
$=2n_{_{\rm He}}$. Having voluntarily chosen to use downstairs chemical 
indices for the currents, we have no option but to use upstairs indices 
for the masses because their corresponding transformation will be of 
contravariant kind (i.e. given by the inverse of the covariant 
transformation matrix) that in this particular illustration will be 
specified by the relations $m^{_{\rm H}}=m^{_{\rm p}}$, 
$m^{_{\rm He}}$ $=2m^{_{\rm p}} + 2m^{_{\rm n}}$. At the cost perhaps of 
obscuring other relevant information, such a change of chemical basis 
(from the atomic reference system to the nuclear reference system) would 
have the advantage of facilitating the exploitation of the empirical fact 
that for many practical applications (particularly in contexts for which 
a Newtonian description is sufficiently accurate) it will be a good 
enough approximation to take $m^{_{\rm n}}\sim m^{_{\rm p}}$ (a relation 
attributable to the corresponding, but still theoretically unexplained, 
approximate inequality between up and down -- but not strange -- quark 
masses). 

Having dealt with the gravitational potential contribution, we now
turn to the kinetic contribution in the decomposition (\ref{50c}) of 
$\Lambda_{_{\rm kin}}$, for which a multiconstituent version can be 
obtained simply by adding up the contributions, as specified by 
(\ref{48c}), of the separate constituents. We thus obtain a prescription
of the form
\be \Lambda_{_{\rm kin}}={_\sum \atop ^{_{\rm X}}}
 n_{_{\rm X}}^{\,\nu}p^{_{\rm X}}_{\, \nu}\, ,\label{64}\fe
in which, as in (\ref{63a}), the sum on the right is covariant with 
respect to spacetime coordinate transformations. However despite its 
neat appearance (but due to the non-linearity in the defining formula 
(\ref{38b}) for relevant kinetic momentum covectors 
$p^{_{\rm X}}_{\, \nu}$) this contribution (\ref{64}) is neither gauge 
invariant (with respect to non-linear Milne or even linear Galilean 
transformations) nor invariant under changes of chemical basis. 

In the manner shown by (\ref{41d}) we can recover gauge invariance of 
the global integrated action perturbation (though not of the local 
unperturbed action density) by combining the kinetic contribution with 
the gravitational potential contribution. This gauge invariance at the 
global perturbation level will evidently be preserved by the addition of  
the extra locally Galilei (and hence a fortiori also Milne) invariant 
term $\Lambda_{_{\rm int}}$ that is needed to give a chemically invariant 
value for the total material contribution $\Lambda_{_{\rm mat}}$, and 
hence also for the complete Lagrangian  (\ref{62}). The variation 
\be \delta \Lambda_{_{\rm mat}}={_\sum \atop ^{_{\rm X}}} 
\mu^{_{\rm X}}_{\,\nu}\,\delta n_{_{\rm X}}^{\, \nu}\, ,\label{65}\fe
of the chemically invariant material Lagrangian will then define a 
chemically contravariant set of material momentum covectors 
$\mu^{_{\rm X}}_{\,\nu}$. These will be decomposible in the form
\be \mu^{_{\rm X}}_{\,\nu}=p^{_{\rm X}}_{\,\nu}+\chi^{_{\rm X}}_{\,\nu}
\, ,\label{65a}\fe
where the internal momentum contributions are defined by the variation 
\be \delta \Lambda_{_{\rm int}}={_\sum \atop ^{_{\rm X}}}
 \chi^{_{\rm X}}_{\,\nu}\,\delta n_{_{\rm X}}^{\, \nu} \, ,\label{65b}\fe
of the internal Lagrangian contribution, whose chemical basis 
dependence endows the momentum contributions $\chi^{_{\rm X}}_{\,\nu}$
with corresponding bad (meaning non contravariant) behavior under
chemical basis transformations so as to cancel the bad behavior of the 
kinetic momentum contributions $p^{_{\rm X}}_{\,\nu}$ in such a way that 
the total (\ref{65a}) is chemically well behaved. However, although 
their chemical transformation behaviour is complicated,  the internal 
momentum contributions $\chi^{_{\rm X}}_{\,\nu}$ have the convenient 
redeeming feature that (unlike the chemically well behaved total 
$\mu^{_{\rm X}}_{\,\nu}$) they are automatically invariant with respect 
to Milne (and therefore a fortiori Galilean) gauge transformations, as a 
consequence of the postulated gauge invariance of the Lagrangian 
contribution $\Lambda_{\rm int}$ itself. In the manner to be derived 
in a subsequent article \cite{CCII}, this gauge invariance entails 
corresponding Noether identities that are expressible as
\be {_\sum \atop ^{_{\rm X}}} t_\mu n_{_{\rm X}}^{\,\mu}
\chi^{_{\rm X}}_{\, \nu}\gamma^{\nu\sigma} = 0\, .\label{66}\fe
and
\be {_\sum \atop ^{_{\rm X}}} n_{_{\rm X}}^{\,[\mu}\gamma^{\sigma]\nu}
\chi^{_{\rm X}}_{\,\nu}=0 \, .\label{66a}\fe

The space projected parts of the internal momentum give rise to the 
effect (of a kind that is familiar in the case of ordinary electron
currents in a metallic conductor) that is known as ``entrainment'' 
whereby the constituent momentum directions may deviate from those of 
the corresponding velocities. However it follows from (\ref{66}) that 
the deviations will cancel out in the total, so that the (gauge dependent 
but chemically invariant) 3-momentum density, i.e. the space projected 
part of the 4-momentum density (\ref{63a}), as defined by 
\be {\mit \Pi}^\mu=\gamma^\mu_{\ \nu}\rho^\nu\, ,\label{67}\fe
will be expressible as a sum over separate material momentum
contributions in the form
\be{\mit \Pi}^\mu={_\sum \atop ^{_{\rm X}}} {\mit\Pi}^{_{\rm X}\mu}\, ,
\label{67a}\fe
in which the individual contributions are given by
\be{\mit\Pi}^{_{\rm X}\mu}= n_{_{\rm X}}\mu^{_{\rm X}}_{\, \nu}
\gamma^{\nu\mu} \, .\label{67b}\fe

In a manner analogous to that by which the ordinary pressure function
$P$ was introduced by (\ref{60}) in the simple perfect fluid case,
it is useful to define a corresponding gauge invariant (and chemically
invariant) generalized pressure function $\Psi$ for the multiconstituent 
case by the specification
\be \Psi=\Lambda_{_{\rm int}} -{_\sum \atop ^{_{\rm X}}} n_{_{\rm X}}^{\,\nu}
\chi^{_{\rm X}}_{\,\nu} \, ,\label{83b}\fe
It can be seen from (\ref{65b}) that the corresponding infinitesimal
variation formula will have the form
\be \delta\Psi= -{_\sum \atop ^{_{\rm X}}} n_{_{\rm X}}^{\,\nu}\,\delta
\chi^{_{\rm X}}_{\,\nu} \, .\label{83c}\fe

Putting the kinetic, internal, and external gravitational potential 
contributions together, we now see that the variation of the complete 
Lagrangian (\ref{62}) will have the (chemically covariant) form
\be \delta\Lambda= {_\sum \atop ^{_{\rm X}}}\pi^{_{\rm X}}_{\,\nu}\,\delta 
n_{_{\rm X}}^{\,\nu} -\rho\,\delta\phi\, ,        \label{68}\fe
while the corresponding variation of the pressure function will
be expressible in the form
\be\delta\Psi=-{_\sum \atop ^{_{\rm X}}}  n_{_{\rm X}}^{\,\nu}\,
\delta \pi^{_{\rm X}}_{\,\nu}-\rho\,\delta\phi\, ,\label{80b}\fe
in which the complete (chemically contravariant) particle momentum 
covectors of the various constituent currents are given by expressions 
of the same form (\ref{50}) as in the single constituent case, namely
\be \pi^{_{\rm X}}_{\,\mu}=\mu^{_{\rm X}}_\mu - m^{_{\rm X}}
\phi\, t_\mu\, .\label{68a}\fe
It is to be observed that since the extra (gravitational) term here is 
purely temporal, it makes no difference to the three momentum, so 
(\ref{67b}) can just as well be written in the form
\be {\mit\Pi}^{_{\rm X}\mu} =  n_{_{\rm X}}\pi^{_{\rm X}}_{\, \nu}
\gamma^{\nu\mu}\, .\label{67c}\fe

As in the case (\ref{50d}) of a simple perfect fluid, we can rewrite the
complete Lagrangian in terms of the momenta $\pi^{_{\rm X}}_{\,\nu}$
and the generalized pressure function $\Psi$ in the form
\be \Lambda={_\sum \atop ^{_{\rm X}}}n_{_{\rm X}}^{\,\nu}
\pi^{_{\rm X}}_{\,\nu}+\Psi\, .\label{80a}\fe
Since, as in the single constituent case, the effect of an infinitesimal 
gauge transformation on the momenta will be given in terms of the relevant 
boost potential $\beta$ by an expression of the simple form
\be\breve {\rm d}\pi^{_{\rm X}}_{\,\mu}=-m^{_{\rm X}}\nabla_{\!\mu}\beta\, ,
\label{63d}\fe
while the other quantities in (\ref{80a}) will remain invariant, it is
evident that the resulting infinitesimal gauge variation of the Lagrangian 
will be given by the formula
\be \breve {\rm d}\Lambda = - {_\sum \atop ^{_{\rm X}}}m^{_{\rm X}}
 n_{_{\rm X}}^{\,\nu}\nabla_{\!\nu}\beta\, .\label{63e}\fe
Subject to the (chemically invariant) restriction that each of the currents 
should be separately conserved, 
\be \nabla_{\!\nu}n_{_{\rm X}}^{\,\nu}=0\, ,\label{63f}\fe
it can be seen to follow that  $\breve {\rm d}\Lambda$ will be expressible
as a pure divergence of exactly the same simple form (\ref{63c}) as in the 
single constituent case. Since it can be seen from (\ref{45c}) that the 
restriction (\ref{63f}) is preserved by the generic flow displacement 
(\ref{45}) it follows that the corresponding variational equations of motion 
that we shall derive below will be gauge independent.

It is to be observed that the condition (\ref{63c}) for the gauge invariance 
of the integrated action will still be satisfied even if the currents do not 
satisfy the separate conservation conditions (\ref{63f}) but are restricted 
only by the single conservation condition (\ref{63b}), that must always be 
satisfied by the total Newtonian mass current (\ref{63a}). However, 
if the currents are not separately conserved, the total mass conservation
condition (\ref{63b}) will not automatically be preserved by a generic
set of independent flow displacements of the form (\ref{45}),
so the variational principle will no longer provide an automatically
gauge invariant set of field equations. 

When each current $n_{_{\rm X}}^{\,\nu}$ is subject to its own independent
displacement $\xi_{_{\rm X}}^{\,\nu}$, the generalization of (\ref{51})
that we finally obtain by substituting (\ref{45b}) in (\ref{68}) will
take the form
\be \delta\Lambda =\nabla_{\!\mu}\big(2{_\sum \atop ^{_{\rm X}}} 
\pi^{_{\rm X}}_{\,\nu} n_{_{\rm X}}^{[\mu}\xi_{_{\rm X}}^{\,\nu]} \big) 
-{_\sum \atop ^{_{\rm X}}} f^{_{\rm X}}_{\,\mu} \xi_{_{\rm X}}^{\,\mu}
-\rho\,\delta\phi\, ,\label{68b}\fe
in which, for each constituent, the covector $f_\mu^{_{\rm X}}$ is 
interpretable as the 4-force density acting on the corresponding current 
$n_{_{\rm X}}^{\,\mu}$, and in which the value of this force density can 
be read out as
\be f_\mu^{_{\rm X}} = 2 n_{_{\rm X}}^{\,\nu}\nabla_{\![\nu}
\pi^{_{\rm X}}{_{\!\mu]}}+\pi^{_{\rm X}}{_{\!\mu}}\nabla_{\!\nu}
n_{_{\rm X}}^{\,\nu}\, .\label{68c}\fe
It is this that must vanish in the strictly conservative case for
which the variational field equations are satisfied.

Whenever the separate conservation conditions (\ref{63f}) actually are 
satisfied, the force density will reduce to the simple form
\be f_\mu^{_{\rm X}} = n_{_{\rm X}}^{\,\nu}
\varpi^{_{\rm X}}{_{\!\nu\mu}}\, ,\label{70}\fe
using the notation
\be \varpi^{_{\rm X}}{_{\!\mu\nu}}=2\nabla_{\![\mu}
\pi^{_{\rm X}}{_{\!\nu]}}\, ,\label{70a}\fe
for the generalized vorticity tensors, which unlike the momenta 
from which they are derived, can be seen from the form of (\ref{63d})
to be gauge invariant,
\be \breve {\rm d} \varpi^{_{\rm X}}{_{\!\mu\nu}}=0\, .\label{70b}\fe
Although the decay rates (\ref{61}) are also gauge invariant,
 \be \breve {\rm d} {\cal D}_{_{\rm X}}=0\, ,\label{70c}\fe
whenever they are non zero, i.e. when the separate conservation 
conditions (\ref{63f}) are not satisfied, the resulting force density
contributions
\be f_\mu^{_{\rm X}} = n_{_{\rm X}}^{\,\nu}
\varpi^{_{\rm X}}{_{\!\nu\mu}}-\pi^{_{\rm X}}{_{\!\mu}}
{\cal D}_{_{\rm X}}\, ,\label{70d}\fe
will no longer be gauge invariant, but will transform according to the rule
\be \breve {\rm d} f_\mu^{_{\rm X}}=m^{_{\rm X}}{\cal D}_{_{\rm X}}
\nabla_{\!\mu} \beta \, .\label{70e}\fe

It can be seen from the general formula (\ref{59k}) that when the separate
current conservation conditions (\ref{63f}) are satisfied, the Lie
derivatives of the vorticities with respect to the corresponding flow 
fields will be given in terms of the corresponding force densities by 
\be \vec n_{_{\rm X}} \Libra \varpi^{_{\rm X}}{_{\!\mu\nu}}
=2\nabla_{\![\mu}f^{_{\rm X}}{_{\!\!\!\!\nu]}}\, .\label{70k} \fe
When the full set of variational field equations is satisfied, so that the 
forces densities $f^{_{\rm X}}_{\mu}$ all vanish, it can be seen that 
each vorticity will be conserved by transport along the corresponding
flow lines, i.e. each constituent will satisfy a Kelvin type conservation 
law of the form (\ref{59l}), namely
\be (\zeta^{_{\rm X}}\vec n_{_{\rm X}}) \Libra \varpi^{_{\rm X}}{_{\!\mu\nu}} 
= 0 \, .\label{70l}\fe
for arbitrary scalar fields $\zeta^{_{\rm X}}$.

As in the  corresponding relativistic case~\cite{CarterKhalatnikov92},
one can go on to generalize the single constituent helicity vector
(\ref{58d}) to a vector valued helicity matrix defined by
\be \eta^{_{\rm XY}\mu}={_1\over^2}\varepsilon^{\mu\nu\rho\sigma}
\pi^{_{\rm X}}{_{\!\nu}}\varpi^{_{\rm Y}}{_{\!\rho\sigma}}
\, .\label{72}\fe
The antisymmetric part of this matrix will be exact in the sense of
having the form of a divergence, 
\be \eta^{_{\rm [XY]}\mu}={_1\over^2}\nabla_{\!\nu}\big(
\varepsilon^{\mu\nu\rho\sigma}\pi^{_{\rm Y}}{_{\!\rho}}
\pi^{_{\rm X}}{_{\!\sigma}}\big)\, ,\label{72b}\fe
of an antisymmetric tensor, with the implication that it will
automatically be closed in the sense that its own divergence
will vanish identically, i.e.
\be \nabla_{\!\mu}\eta^{_{\rm [XY]}\mu}=0\, .\label{72c}\fe
It follows that if a constituent with label $_{\rm X}$ say is
characterized by the property of irrotationality, meaning that
$\varpi^{_{\rm X}}{_{\!\mu\nu}}$ vanishes, and more particularly, as in 
(\ref{59i}), in the case of a constituent that is superfluid, and thus
characterized (on a mesoscopic scale) by a momentum covector that is the 
gradient of a corresponding phase scalar $\varphi^{_{\rm X }}$,  so that 
for any other label value $_{\rm Y}$ say, the corresponding helicity 
matrix component $\eta^{_{\rm YX}\mu}$ will also vanish, i.e. 
$\eta^{_{\rm YX}\mu}=0$, then the corresponding transposed component will 
automatically be conserved, i.e. we shall have 
\be\pi^{_{\rm X}}{_{\!\mu}}=\nabla_{\!\mu}\varphi^{_{\rm X }}\hskip 0.6 cm
\Rightarrow \hskip 0.6 cm \varpi^{_{\rm X}}{_{\!\mu\nu}}=0 \hskip 0.6 cm
\Rightarrow \hskip 0.6 cm \nabla_{\!\mu}\eta^{_{\rm XY}\mu}=0\, .
\label{73}\fe 

Regardless of any irrotationality constraint that may be satisfied, it
can be seen -- for the same reason as in the single constituent case 
characterized by (\ref{59f}) -- that the vanishing of the force density 
covector $f_\mu^{_{\rm X}}$ in (\ref{70}) will always be sufficient to 
ensure that the divergence of the corresponding diagonal component of the 
helicity matrix will vanish, i.e. that the constituent under consideration
will be subject to a helicity conservation 
law~\cite{Peradzynski90,CarterKhalatnikov92} having the form
\be \nabla_{\!\mu}\eta^{_{\rm XX}\mu}=0\, .\label{74}\fe

\bigskip
{\bf Acknowledgements}
\medskip

The authors wish to thank Silvano Bonazzola, Eric Gourgoulhon, Keith
Moffat and Reinhard Prix for instructive conversations.

\vfill\eject

\bigskip
{\bf Appendix: helicity current in the traditional 3+1 formalism}
\medskip

The efficacity of 4-dimensionally covariant treatment is clearly
demonstrated by the ease with which the foregoing helicity
conservation laws have been obtained as almost obvious
consequences of the dynamical equation. This contrasts with
treatment using a 3+1 space versus time decomposition, in which an 
equivalent derivation of the helicity conservation law~\cite{Moffat69}
requires much greater algebraic effort and ingenuity, as shown in this
appendix.

The time and space components of the helicity current are found from
 (\ref{58d}) and the specification (\ref{04}) to be expressible as
\be \eta^{_0} = -\varepsilon^{ijk}\pi_i\nabla_{\!j}\pi_k \, ,\fe
\be \eta^i = \pi_{_0} \varepsilon^{ijk}\nabla_{\!j}\pi_k-\varepsilon^{ijk}
\pi_j\nabla_{\!_0}\pi_k+\varepsilon^{ijk}\pi_j\nabla_{\!k}\pi_{_0}\, . \fe

In the following -- as is usual within an Aristotelian - Cartesian framework.
-- the space indices will be replaced by the familiar arrow notation. We also 
introduce the cross-product $(\phi \times \varphi)^i$ between
two forms $\phi_i$ and $\varphi_i$ as the contravariant vector 
$\varepsilon^{ijk}\phi_j\varphi_k$. The curl and 3-divergence of a 
vector $\vec V$ are defined respectively as 
$(\vec{\rm curl}\, \vec V)^i=\varepsilon^{ijk}\nabla_{\! j} V_k$ and 
${\rm div}\, \vec V = \nabla_{\!i} V^i$. The time derivative will be
written as $\nabla_{\!_0} = \partial_t $.

With this notation, the helicity 4-vector will be given by
\be \eta^{_0} = -\vec \pi.\vec{\rm curl}\, \vec \pi\, ,\fe
\be \vec \eta = \pi_{_0}\, \vec {\rm curl}\, \pi-\vec\pi \times \partial_t
\vec \pi+\vec\pi \times \vec\nabla\pi_{_0}\, . \fe

Considering the example of the single perfect fluid model, the components 
of the 4-momentum covector (\ref{50}) are $\pi_{_0}=-{\cal E}$ and
$\vec\pi= m\vec v$, in which we have introduced the total particle energy 
${\cal E} ={_1\over^2}m v^2+m\phi+\chi$. Hence the helicity current reduces to
\be \eta^{_0} = -2m^2 \vec \omega.\vec v\, ,\fe
\be \vec \eta =  -2m{\cal E} \vec \omega+\vec\nabla {\cal E} 
\times m \vec v+m^2 \partial_t \vec v \times\vec v\fe
in terms of the kinematic local angular velocity $\vec \omega 
= {_1\over^2}\vec{\rm curl}\,\vec v$ (${\rm div}\, \vec \omega = 0$).

The conservation law of this helicity current will now be derived within 
the 3+1 spacetime decomposition, starting from Euler's equation in the 
form~\cite{Landau}
\be m\partial_t \vec v + \vec \nabla {\cal E} + 2m\vec \omega \times 
\vec v = \vec  0\, ,\fe
whose contraction with $m\vec\omega$ yields
\be m^2 \vec\omega.\partial_t\vec v + m\vec\omega.\vec \nabla{\cal E} 
= 0 \, .\fe
Taking the curl of the Euler equation followed by 
the dot product with the velocity gives
\be m^2 \vec v.\partial_t \vec \omega+m^2\vec v.\vec {\rm curl} 
(\vec \omega \times \vec v)=0\, .\fe
Adding the two equalities and simplifying, we obtain
\be 2m^2\partial_t(\vec\omega.\vec v)+2 {\rm div} (m\vec \omega 
{\cal E})+2m^2 (\vec v.\vec \omega) {\rm div}\,\vec v +
2m^2\vec v.(\vec v.\vec \nabla \vec \omega)-2m^2\vec v.(\vec \omega.
\vec \nabla \vec v) = 0\, . \label{180}\fe
Taking the cross-product of the Euler equation with the velocity, the divergence and
using the standard identity 
\be \vec \nabla (\vec W.\vec V) = \vec W\times \vec {\rm curl}\,\vec V
+\vec V\times\vec{\rm curl}\,\vec W+\vec V.\vec \nabla \vec W
+\vec W.\vec \nabla \vec V\, , \fe
we obtain 
$$ {\rm div}(m^2 \vec v\times\partial_t\vec v+
m \vec v\times\vec \nabla{\cal E})+8m^2\vec \omega.(\vec v\times\vec \omega)
+4m^2\vec \omega.(\vec v.\vec \nabla \vec v)$$ 
\be -2m^2(\vec v.\vec \omega){\rm div}\,\vec v -2m^2\vec v. 
(\vec v\times\vec {\rm curl}\, \vec \omega)
-2m^2 \vec v.(\vec v\vec \nabla \vec \omega)
-2m^2\vec v.(\vec \omega.\vec\nabla \vec v) = 0\, .\fe
Combining this with (\ref{180}), one obtains the space-time decomposition 
of (\ref{59f}), namely the local version of the original Moreau-Moffat 
helicity conservation law~\cite{Moreau61,Moffat69}, in the form 
\be 2m^2\partial_t(\vec\omega.\vec v)={\rm div} 
\biggl(-2m\vec \omega {\cal E}+m^2\partial_t\vec v\times \vec v
+m(\vec\nabla{\cal E})\times\vec v\biggr) \, .\fe

\end{document}